\def\year{2019}\relax
\documentclass[letterpaper]{article} 
\usepackage{aaai19}  
\usepackage{times}  
\usepackage{helvet} 
\usepackage{courier}  
\usepackage[hyphens]{url}  
\usepackage{graphicx} 
\def\UrlFont{\rm}  
\usepackage{graphicx}  
\usepackage{colortbl}
\usepackage{booktabs} 
\usepackage{footmisc}
\usepackage{multirow}
\usepackage{todonotes}

\usepackage{wrapfig}

\usepackage{subfig}
\usepackage{color}
\usepackage{enumitem}

\nocopyright

\title{Competing Topic Naming Conventions in Quora: Predicting Appropriate Topic Merges and Winning Topics from Millions of Topic Pairs}
\author{Binny Mathew, Suman Kalyan Maity,\textsuperscript{\rm 1} Pawan Goyal, Animesh Mukherjee\\
Indian Institute of Technology(IIT), Kharagpur\\
\textsuperscript{\rm 1}Kellogg School of Management, Northwestern University\\
binnymathew@iitkgp.ac.in, suman.maity@kellogg.northwestern.edu, \{pawang, animeshm\}@cse.iitkgp.ac.in
}

\begin{document}

\maketitle

\begin{abstract}

Quora is a popular Q\&A site which provides users with the ability to tag questions with multiple relevant topics which helps to attract quality answers. These topics are not predefined but user-defined conventions and it is not so rare to have multiple such conventions present in the Quora ecosystem describing exactly the same concept. In almost all such cases, users (or Quora moderators) manually merge the topic pair into one of the either topics, thus selecting one of the competing conventions. An important application for the site therefore is to identify such competing conventions early enough that should merge in future. In this paper, we propose a \textit{two-step approach that uniquely combines the anomaly detection and the supervised classification frameworks} to predict whether two topics from among millions of topic pairs are indeed competing conventions, and should merge, achieving an F-score of 0.711. We also develop a model to predict the direction of the topic merge, i.e., the winning convention,  achieving an F-score of 0.898. Our system is also able to predict $\sim 25$\% of the correct case of merges within the first month of the merge and $\sim 40$\% of the cases within a year. This is an encouraging result since Quora users on average take 936 days to identify such a correct merge. Human judgment experiments show that our system is able to predict almost all the correct cases that humans can predict plus 37.24\% correct cases which the humans are not able to identify at all.
\end{abstract}

\section{Introduction}

Conventions form an important part of our daily life. They affect our basic perceptions~\cite{asch1955opinions} as well as guide everyday
behaviors of the general public~\cite{ajzen1970prediction}.  With the emergence of social media, conventions started being used in online space. Twitter, for example, follows conventions like "\#" used for specifying keywords, "@" for specifying mentions, and ``RT'' which is an acronym for retweet. None of these were defined by Twitter, but gradually became conventions~\cite{kooti2012emergence} due to their usability and wide acceptance among the users. Similar competing conventions has been observed in other places as well~\cite{rotabi2017competition,coscia2013competition}.
In this paper, we investigate the competition among the topic naming conventions prevalent in Quora.

\begin{figure*}[ht]
	\centering
	\subfloat[Cumulative growth of questions posted on Quora]{%
		\includegraphics[width=.30\textwidth]{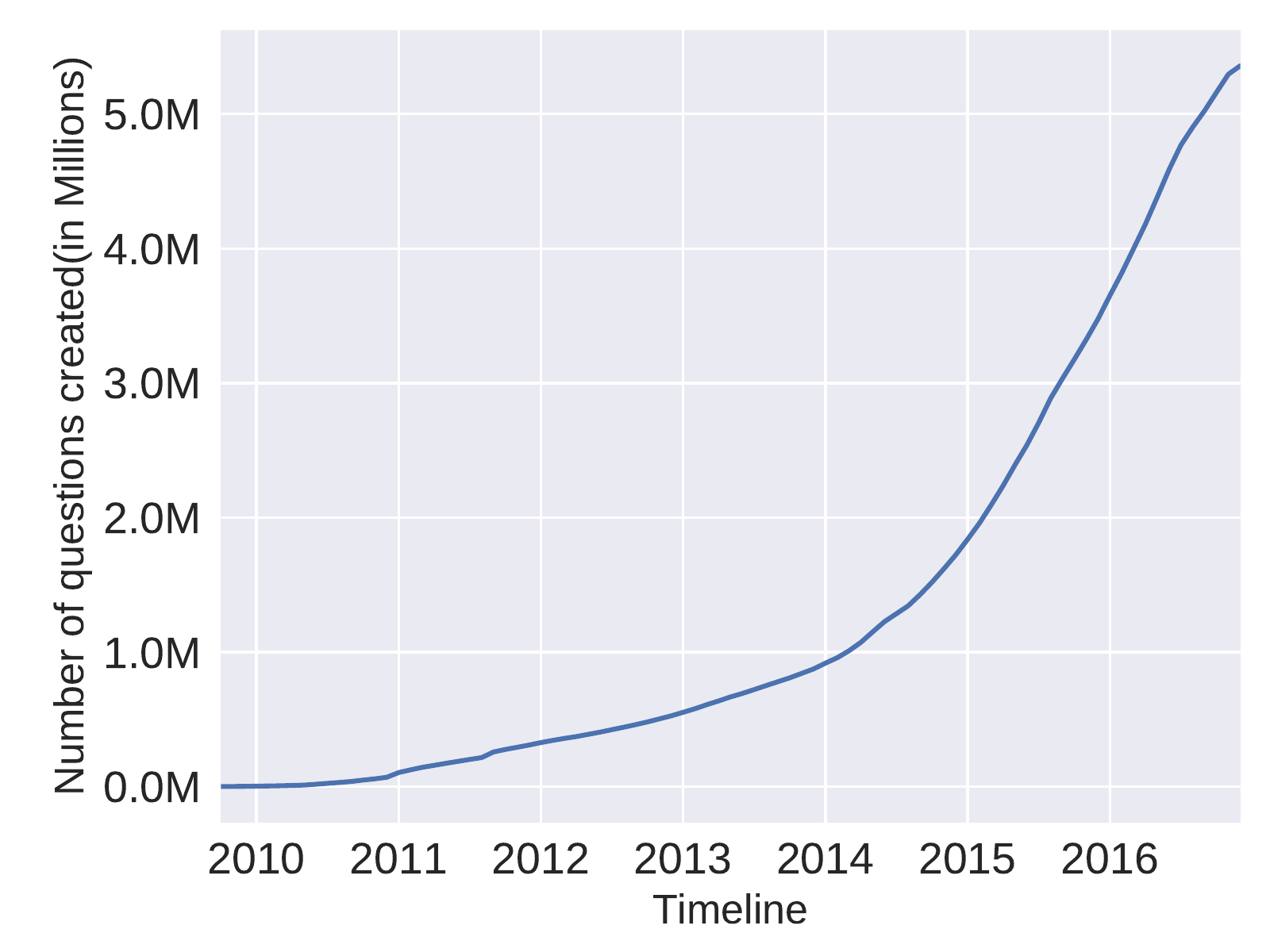} \label{fig:figure1}}\hfill
	\subfloat[Cumulative growth of topics created on Quora]{%
		\includegraphics[width=.30\textwidth]{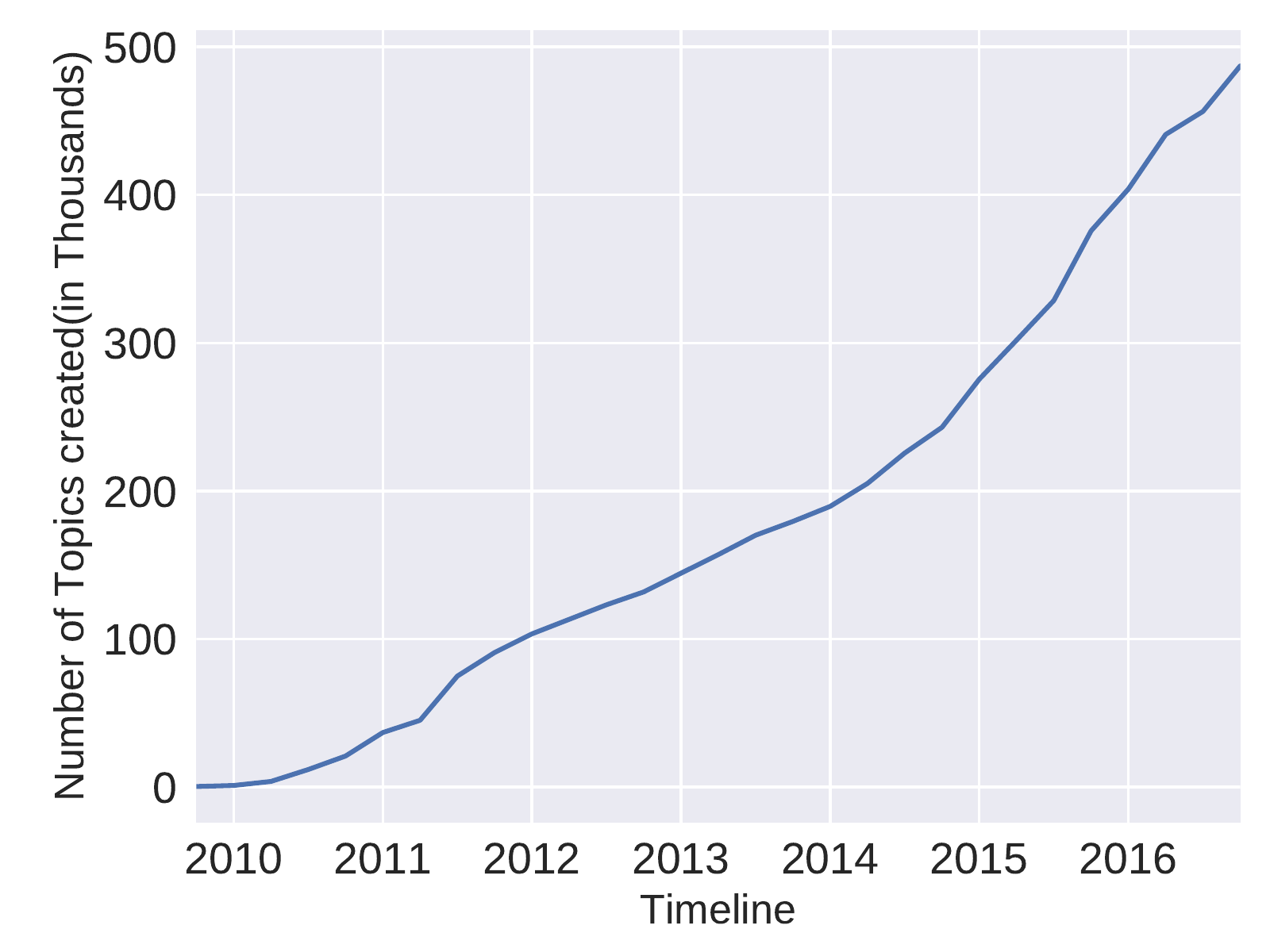} \label{fig:figure2}}\hfill
	\subfloat[The Cumulative growth of Topic Merge and Topic Unmerge in Quora]{%
		\includegraphics[width=.30\textwidth, height=0.164\textheight]{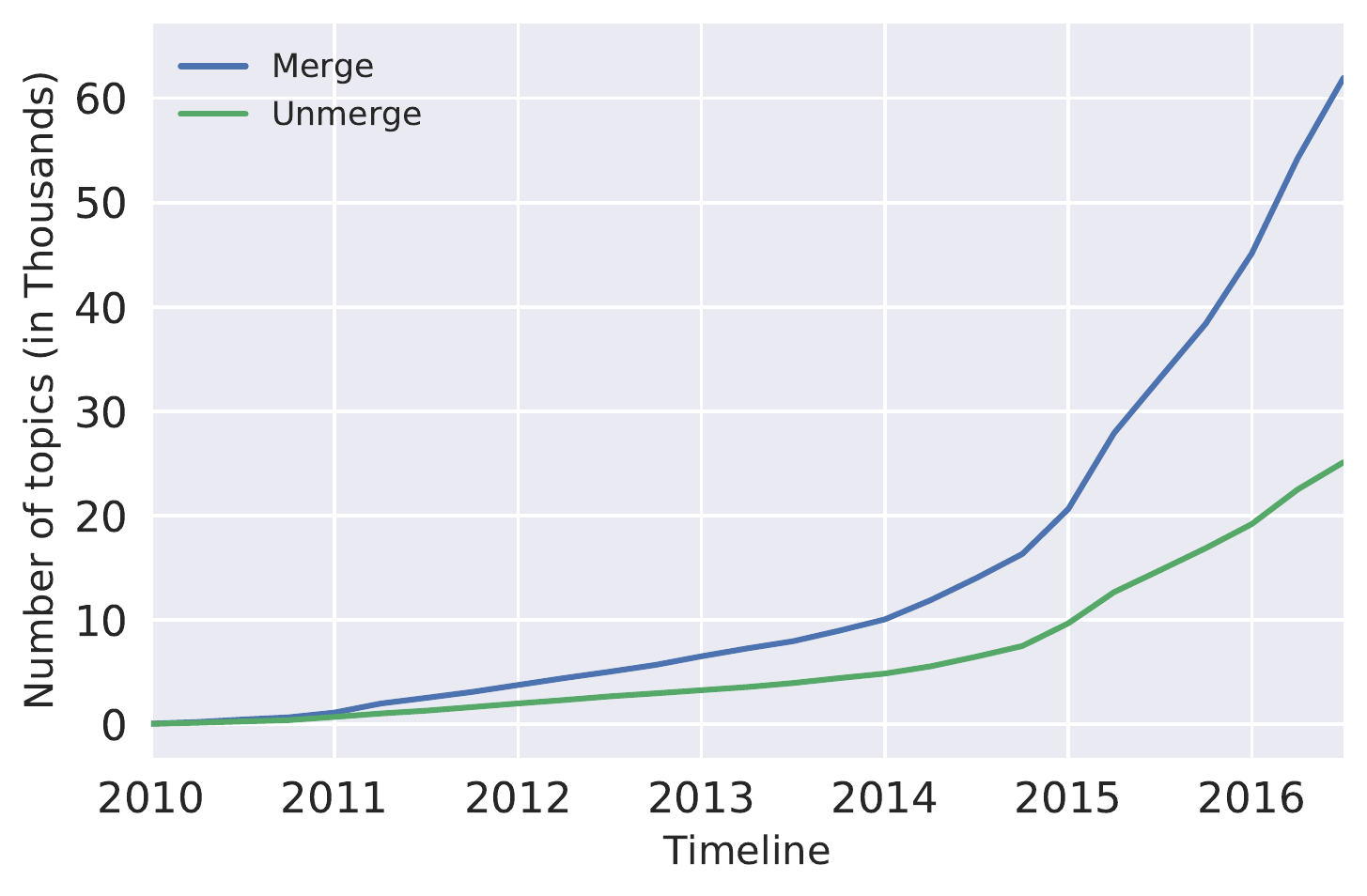} \label{fig:merge_unmerge_timeline}}\hfill
	
	\caption{Question and topic growth on Quora.}

\end{figure*}

Quora is a popular Q\&A site with $\sim$100
million\footnote{https://www.quora.com/How-many-people-use-Quora-7/answer/Adam-DAngelo}
monthly unique visitors. 
As of April 2017, 13 million questions~\cite{wang2013wisdom} have been
posted on Quora. We have obtained $\sim 5.4$ million questions which
is a significant proportion of all the Quora questions and to the best
of our knowledge, is the largest Quora question base.
Figure~\ref{fig:figure1} shows the cumulative growth of the number of questions posted on Quora. Each of these questions are assigned topics that allows to categorize these questions. These topics are defined by the user according to his/her convenience. Figure~\ref{fig:figure2} shows the cumulative growth of the number of topics being created on Quora.

The choice of Quora topics for this study is motivated by the crucial roles these topics play in Quora's ecosystem of knowledge. For example, people follow topics to indicate their interests, which helps Quora to show them content that they should find valuable and engaging. Meanwhile, when people add questions, they tag these with relevant topics so that the question can be channeled to answerers who have relevant expertise, as well as to those who want to learn more about the matter. To complement this, experts can identify topics in which they have specialist knowledge and about which they can provide compelling answers. In the ideal case, there should exist only one topic for each concept. But in several cases, users tend to create multiple topics to represent the same concept. Quora deals with this issue by allowing users to merge topics which represent the same concept.

\subsection{Motivation}
Since topics on Quora are user-defined conventions to express certain
concepts, topic merging on Quora can be considered as a competition
among conventions in which the topic names which are merged can be
considered to be conventions and the competition between these
conventions determines the winner topic name which can better represent the underlying concept. These conventions might be created at different
time periods and a newly created topic name may represent the concept better than the existing
topic name. In such a case, the users can replace the old convention with
the newly formed convention. We find that in case of Quora, 33\% of the times
an existing convention gets replaced by a new convention.

The topics on Quora allows the users to group questions of similar interest. If multiple topics are allowed to represent the same concept, it would degrade the quality of the questions and possibly hamper correct routing of the questions to the correct experts who actually are able to answer the questions. In our dataset, we observe that a Quora user takes on average 936 days to merge duplicate topics from its date of creation. This means that whatever question is posted with the new duplicate topic name, it would most likely miss the actual topic audience. This in turn reduces the quality of the answers received to these questions. In this paper, we propose to characterize the competing topic naming conventions on Quora, and develop various features that can effectively indicate, at an early stage, if a pair of topics that are competing, would merge in future. This would allow the community to merge topics at an early stage of the topic evolution, in turn, promoting early and appropriate knowledge aggregation. Using our model, we are able to predict around 25\% of the merge cases in the first month of the topic creation itself. This would allow the duplicate topics to be merged at an early stage itself, thereby, increasing the quality of the answers received.

\subsection{Research objectives and contributions}
In this paper, we study the phenomena of topic merging in Quora
as competing conventions, and investigate various factors influencing
this phenomena. Toward this objective, we make the following
contributions: 
\begin{itemize}[noitemsep,nolistsep,leftmargin=*]
	\item We obtain a massive dataset and investigate the detailed characteristics of the merging topics. We also introduce the Quora topic ontology - a well defined hierarchy of topics in Quora with each topic linked to its parent topic and child topics.
	\item We propose a two-step approach which \textit{uniquely combines the anomaly detection with the supervised classification framework} to automatically predict whether a topic pair is competing and would get merged in future, achieving an F-score of $\sim 0.711$ for predicting true merges from a million test instances. We propose another
	classification model to predict the winner of the competition
	and achieve an F-score of 0.898.
    \item As a direct application, we also launch an early prediction scheme, whereby, we show that we are able to predict $\sim 25$\% of the correct case of merges within the first month of the merge and $\sim 40$\% of the cases within a year. This is especially encouraging given that users on average, as per our dataset, take as high as 936 days to identify a correct merge. This early prediction scheme can be easily integrated to the Quora system thus enabling users and moderators to identify and correctly merge topics much early in time. 
	\item  We further perform an experiment with human subjects to
	identify how well humans can predict whether two topics should
	merge or not, as well as, the direction of topic merge. We observe
	that our system is able to predict almost all the cases which the humans
	are able to correctly predict; further, our system is able to
	predict another 37.24\% correct cases which the humans are not able
	to identify at all. We perform a fine-grained analysis based on the participants' familiarity with the topics, which indicates that only participants very familiar with the topics were able to outperform our system. In addition, humans perform very poorly in
	predicting the direction of the topic merge which our system is
	able to predict highly accurately. 
\end{itemize}

Our study can be summarized by two very unique and interesting observations about the competing topic naming conventions. First, the content of the questions tied to the competing topics and the distance of the topics themselves in the Quora topic ontology together serve as strong indicators of whether the topic pairs are competing and would merge in future. Second, the winning topic among the competing pair is best determined by factors like number of characters/words in the topic name, date of creation of the topic and the number of questions/answers tied to the topic.

The organization of the paper is as follows. Section II provides an overview of the previous research related to the merging. In Section III, we provide several statistics for the dataset as well as the filters applied to generate the dataset. Section IV provides detailed description about the various properties of the merge and non-merge topics. Section V provides the prediction framework for the topic merging.
In Section VI, we perform human judgement experiments to evaluate the human performance.
Section VII performs the correspondence analysis in which we compare the human judgement with our prediction framework.
In Section VIII, we provide several insights about the topic merging process. Section IX concludes the paper along with the future works.

\section{Related works}
Social Q\&A sites allow users to post questions and get quality answers from the community. Sites like Yahoo! Answers, Baidu Knows, Stack Overflow and Quora have been a subject of active 
research~\cite{patil2016detecting,PRA2:PRA2145052010042,maity2015analysis,shen2015,Zhang:2014:QRH:2661829.2661908,mathew2018deep,wang2013wisdom,paul2012authoritative,pal2012evolution,li2010routing,yang2009seeking,adamic2008knowledge}. Patil et al.~\cite{patil2016detecting} study experts and non-
experts in Quora and develop statistical models to automatically detect experts. Maity et al.~\cite{maity2015analysis} analyze the dynamics of topical growth over time and propose a regression model to predict the popularity of the topics. The studies done in~\cite{adamic2008knowledge,li2010routing,pal2012evolution} focus on ranking the users from expertise measures based on users' history and activities. Another direction of research focuses on the quality of the user generated content in Q\&A sites that includes quality of questions~\cite{anderson2012discovering,li2012analyzing} and quality of answers~\cite{adamic2008knowledge,shah2010evaluating,tausczik2011predicting}.

\subsection{Conventions}

Several studies have been conducted to study the emergence of
conventions. One group of such studies is based on small scale experimental work. In Wilkes-Gibbs and Clark~\cite{wilkes1992coordinating} for instance, the authors asked the participants to develop short-hand verbals which allowed them to communicate more efficiently and complete the tasks quickly. In another work~\cite{selten2007emergence}, the authors investigate in a series of laboratory experiments how costs and benefits of linguistic communication affect the emergence of simple languages in a coordination task when no common language is available in the beginning. In both these studies, the need for efficient coordination gave rise to conventions.

Social conventions and their emergence is a fairly old area of study ~\cite{david1969convention,edna1977emergence}. According to Edna~\cite{edna1977emergence}, social
conventions correspond to special type of norms related to coordination problems, that is, conventions are ``those regularities of behavior which owe either their origin or their durability to their being
solutions to recurrent (or continuous) co-ordination problems, and
which, with time, turn normative''. One of the interesting ways to
study the dynamics of conventions using real data is through social media~\cite{yee2007unbearable,friedman2007spatial,kooti2012emergence,coscia2013competition,kooti2012predicting}. In Kooti et al.~\cite{kooti2012emergence} the authors study the emergence of retweet convention in Twitter. They perform in-depth study of the conventions and provide valuable insights. Further, in Kooti et al.~\cite{kooti2012predicting} the authors perform classification analysis and demonstrate that the date of adoption and the number of exposures are particularly important in the adoption process, while personal features (such as the number of
followers and join date) and the number of adopter friends have less discriminative power in predicting convention adoptions. Our results are in line with these findings as we find that the date of creation of a convention has a strong influence in deciding which convention will win. In Rotabi et al.~\cite{rotabi2017competition}, the authors study conventions that vary even as the underlying meaning remains constant by tracking the spread of macros and other author-defined conventions in the e-print arXiv over 24 years. They found that the interaction among co-authors over time plays a crucial role in the selection of conventions. Young authors tend to win the fight which have low visibility consequences, while older authors tend to win fights that produce highly visible consequences.

\subsection{Blending and Compounding}
Lexical blending is a linguistic phenomenon in which a word is formed by fusing two or more words into one another (e.g., biopic = biography + picture, pulsar = pulse + quasar). This linguistic form of word reduction has been studied widely~\cite{brdar2008marginality,carter2002prosodic,connolly2013innovation,cook2012using,cook2010automatically,gaskell1999ambiguity,leturgie2012dictionaries,Medler2004ProcessingAW,renner2013cross}. 

Another closely related phenomena is lexical compounding. Few studies have been conducted on lexical compounding in English~\cite{badecker2001lexical,giegerichCompoundingLexicalism} and other languages like Italian, French, German,
Spanish, Chinese etc.~\cite{arcodia2007chinese,gaeta2009composita,lee2010qualia,packardjl2010morphology,villoing2012french}. Hashtag compounding is a linguistic process in which two separate hashtags coalesce to form a single hashtag. Mitra et al.~\cite{mitra2015automatic} gave an approach to identify word sense changes in text media across timescales. In this work, they talk about word sense merging, in which two senses of the same word merge into a single prevalent sense. Hashtag compounding is a linguistic process in which two separate hashtags coalesce to form a single hashtag. Maity et al.~\cite{MaitySM16} studied various socio-linguistic properties responsible for hashtag compound formation and proposed a model that can identify if a compounded hashtag will become more popular than its individual constituent hashtags.

\subsection{The present work}
The present work significantly differs from the above lines of research. As outlined in the introduction we model the problem of topic merging as a competition of conventions motivated by a few previous studies~\cite{kooti2012emergence,rotabi2017competition}. The major difference of our work  as compared to the work done in the conventions is in using the topic hierarchy information for the competition of conventions. In case of blending and compounding, the two words join to form a new word, but in our problem, one topic is actually merging into another. So no new topic is generated as opposed to blending and compounding.
For the first time, we show how the competition of conventions is driven by the content of the questions corresponding to the two candidate topics and their distance in the \textit{topic ontology}. As an added novelty, we also develop a two-step approach to predict merges as well as an approach to predict the directions of merge.

\begin{figure*}[tb]
	\centering
	\subfloat[Topic merge.]{%
		\includegraphics[width=.32\textwidth]{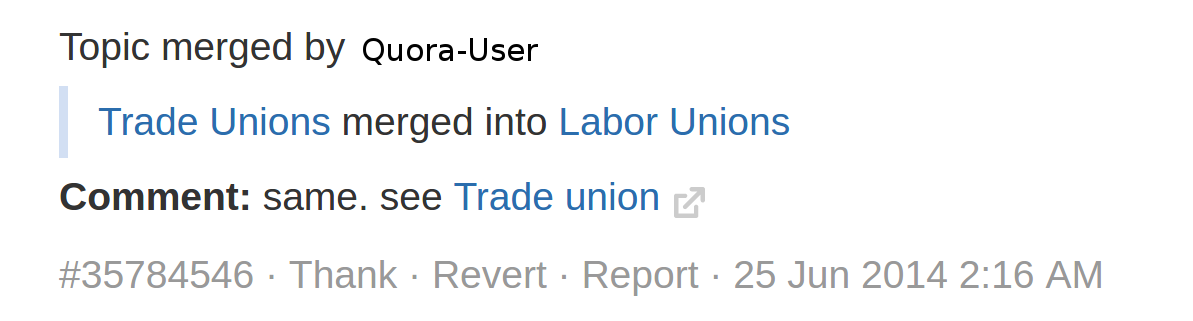} \label{fig:figure3}}\hfill
	\subfloat[Topic unmerge.]{%
		\includegraphics[width=.32\textwidth]{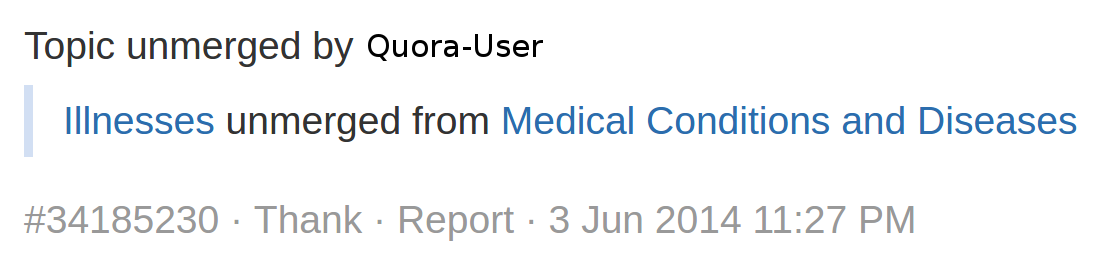} \label{fig:figure4}}\hfill
	\subfloat[Topic parent addition.]{%
		\includegraphics[width=.32\textwidth]{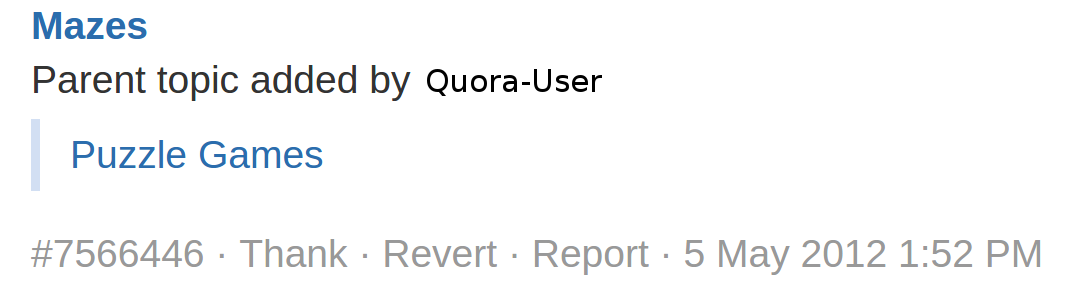} \label{fig:topic_parent_add}}\hfill
	\caption{Topic merge, unmerge and parent addition example.}
\end{figure*}

\begin{table*}[ht]
	\centering
	\scriptsize
	\begin{tabular}{|l| p{1.97cm} p{2.7cm} | p{2.6cm} p{2.6cm} | p{1.90cm} p{1.90cm} |} 
		\hline
		&\textbf{invented-languages}&\textbf{artificial-languages}&\textbf{fake-currency}&\textbf{counterfeit-money}&\textbf{stoned}&\textbf{getting-high} \\\hline
		Question$_1$ &Why are some people so obsessed with the Elvish language (LOTR)?&What are the advantages and disadvantages of creating a language?&How can fake currency affect the economy of a country?&Is it possible to withdraw counterfeit money from a bank machine?&Is it true that you are smarter when you are stoned?&How does smoking get you high?\\
		
		Question$_2$ & Is it possible to create a new language that is not subject to interpretation?&Should someone learn a language or create their own?&What is the punishment for printing fake currency in India?&What are the consequences of counterfeiting money?&How does it feel to get stoned?&Why do people love smoking up and getting high?\\
		
		Question$_3$ & Can Tolkien's languages be studied?&How many languages are there in LOTR?&Can 3D printing be used to make fake currency coins?&How fake notes are circulated in India?&What were the funniest things you said while high?&What do you do when you feel high?\\
		\hline
	\end{tabular}
	\caption{Example questions for competing topics: The topic on the left is merged into the topic on right.}
	~\label{tab:tab_merge_examples}

\end{table*}

\section{Dataset description}

In the ideal situation, each concept in Quora should have only one topic to describe
it. However, this is rarely the case. As topics are not predefined, there can exist multiple topics which correspond to the same concept. Consider the case where the concept
`Euthanasia' is represented by the two topic naming conventions: `Euthanasia' and
`Mercy-Killing'. Having multiple topics for the same concept has
several disadvantages. If someone posts a question on a topic related
to `Euthanasia' and does not use the topic `Mercy-Killing' to tag the
question, then the users who have subscribed to 'Mercy-Killing' will
remain unaware of the question posted. This could reduce the number of
potential quality answers.

To deal with such issues, Quora allows users to merge two topics into one. Figure~\ref{fig:figure3} shows an example where a user merges the topic `Trade Union' (source topic) into `Labor Unions' (destination topic). Topics should only be merged when all of the intents of the two topics under consideration are the same. On the merging of topics, people who follow the source topic and the questions that are tagged under the source topic migrate over to the destination topic and the experience and content is aggregated. Thus, all the questions in `Trade Union' will be merged into `Labor Unions'. Further, all the `Trade Union' tags would be changed to `Labor Unions'. Table~\ref{tab:tab_merge_examples} shows some examples of questions and topics for the merging. The competition between topic naming conventions ensures that the emergent topic is the best representative of the underlying convention as per the wisdom of the Quora user crowd.

Just like two topics can be merged to form a single topic, two merged topics can also be `unmerged' to the original topics back again. This happens when the two merged topics do not represent the same concept as per the observation of the Quora users (or moderators). For example, Figure~\ref{fig:figure4} shows a case in which some user merged the topics `Illness' and `Medical Conditions and Diseases'. As these represent two different concepts, they were later unmerged. Figure~\ref{fig:merge_unmerge_timeline} shows the cumulative growth of the total number of merge and unmerge cases in Quora which seems to be always increasing.

We obtain the Quora dataset from the authors of~\cite{maity2015analysis} and then built up on it to have a massive set of 5.4 million questions and 488,122 topics (along with the topic logs).
In addition to the above dataset, we have also used the Quora topic ontology.

\noindent\textbf{Quora topic ontology:} Quora has an extensive set of topics that is constantly growing and expanding. 
To organize the huge set of topics in Quora, the site managers use parent-child relationships between topics. Figure~\ref{fig:topic_parent_add} shows user adding the topic `Puzzle Games' as the parent of `Mazes'. This topic hierarchy forms a Directed Acyclic Graph\footnote{https://www.quora.com/Are-Quora-topic-hierarchies-a-directed-acyclic-graph}. Each topic can have multiple parents as well as multiple child topics. The root topic of the topic hierarchy graph is ``Major-Topics'' and we can navigate the topic graph down from this topic. 
We extract the topics' parent-child information from the topic logs. Using the parent-child relation, we construct the Quora topic ontology graph. Table~\ref{tab:tab1} shows some of the properties of this ontology.

\begin{table}[htb]
	\centering
	\begin{tabular}{l l} 
		Property & Value \\ \hline\hline
		Number of connected components & 3807  \\ 
		Size of the largest component & 236781  \\
		Number of singletons & 237096 \\
		Depth of the ontology & 17  \\
		Average degree of the graph & 1.36  \\ 
		Average degree of largest component & 2.72\\ 
		\hline
	\end{tabular}
	\caption{Properties of the Quora topic ontology.}
	~\label{tab:tab1}
\end{table}

\subsection{Filters applied}
We use the topic log of 488,122 topics to extract the merge and unmerge information. We find 65,231 cases of topic merges and 38,502 topic unmerges.

\noindent\textbf{Merge:} We apply several filters to the topic merge
pairs. Initially, we remove `trivial' merges, i.e., those merges which are minor lexical variants (e.g., plural
forms). To take care of such cases, we apply the standard Jaro-Winkler similarity
on the topic names and all topic pairs with similarity more than 0.8 were removed. After this, we remove all the merge pairs which were abbreviations like `ICC' which essentially merged into `International Cricket Council'. One should note that if a topic gets merged into another topic, all the questions of the former (source) topic change their tags to the latter (destination) one, so we would not be able to get the questions of the two distinct topics that were being merged. To tackle this problem, we separately re-crawl all the questions for the source and the destination topics by making a call to utility \textit{https://www.quora.com/topic/[source/destination topic name]/all\_questions}\footnote{Example: If the source topic name is `ICC', we can make a call to \textit{https://www.quora.com/topic/ICC/all\_questions} to get all the questions under the topic `ICC'.}. Then we obtain the set of questions that were posted before the merge for both the source and the destination topic. We then filter out those topics that did not have any questions tagged with them. After applying all the above filters, we obtain 2829 merge pairs.

\noindent\textbf{Non-merge:} In order to build a strong competitive negative set to contrast the behavior of the actual merge cases, we define the `non-merge' class. In this class we consider topic unmerge pairs (which should be typically hard to distinguish from the true merges) and the topic pairs in the one-hop neighborhood of the actual merging pairs on the Quora ontology graph (i.e., conceptually close topics to the merging pair but did not themselves undergo any merge). We apply the same filters that we had applied to the merge pairs. Finally, we obtain 11,648 topic neighbor pairs and 2,421 topic unmerge pairs.

\section{Properties of Merge \& non-merge topics}

In this section, we first look into the various reason for topic merging to occur in Quora and then study various characteristic properties that distinguish the actual topic merges from the non-merge cases.
\subsection{Reasons for Topic Merging}
There are several reasons for the existence of competing conventions in the topic names. If one topic name is just the plural form of another topic name, then usually the singular topic name is merged into the plural topic name. For example, the topic name `Master-Mind' is merged into `Master-Minds'. Another reason for competing competing conventions is the use of abbreviations. Topic `ISBN' is merged into the topic `International-Standard-Book-Number'. Some topics which represent a company could have been merged into another topic because either the company had changed its name or it was acquired by another organization. The topic `10gen' was merged into `MongoDB-company' as the company had changed its name. There can be many other reasons for merge. Topic `Ebonics' was merged into `African-American-Vernacular-English' as it represented a better topic name for people to understand.

\subsection{Characteristics of Topic Merging}
We carry out a detailed analysis on the 2,829 topic merge pairs and 14,069 ($11,648$+$2,421$) topic non-merge pairs. 

\subsubsection{Question content features} We derive various properties from the question texts and observe the differences between the topic merge and non-merge pairs.

\noindent\textit{\textbf{n-gram overlap of question texts}}: Question text overlaps between two topics are potential indicators of topic merging. To find out if there are question text overlaps, we extract 1, 2, 3 and 4-grams from the questions belonging to the topic pairs. For each n-gram, we compute the unweighted\footnote{https://en.wikipedia.org/wiki/Overlap\_coefficient} and weighted version of the overlap coefficient.

\begin{figure*}[!htp]
	\centering
	\subfloat[n-gram overlap of question words.]{%
		\includegraphics[width=.33\textwidth]{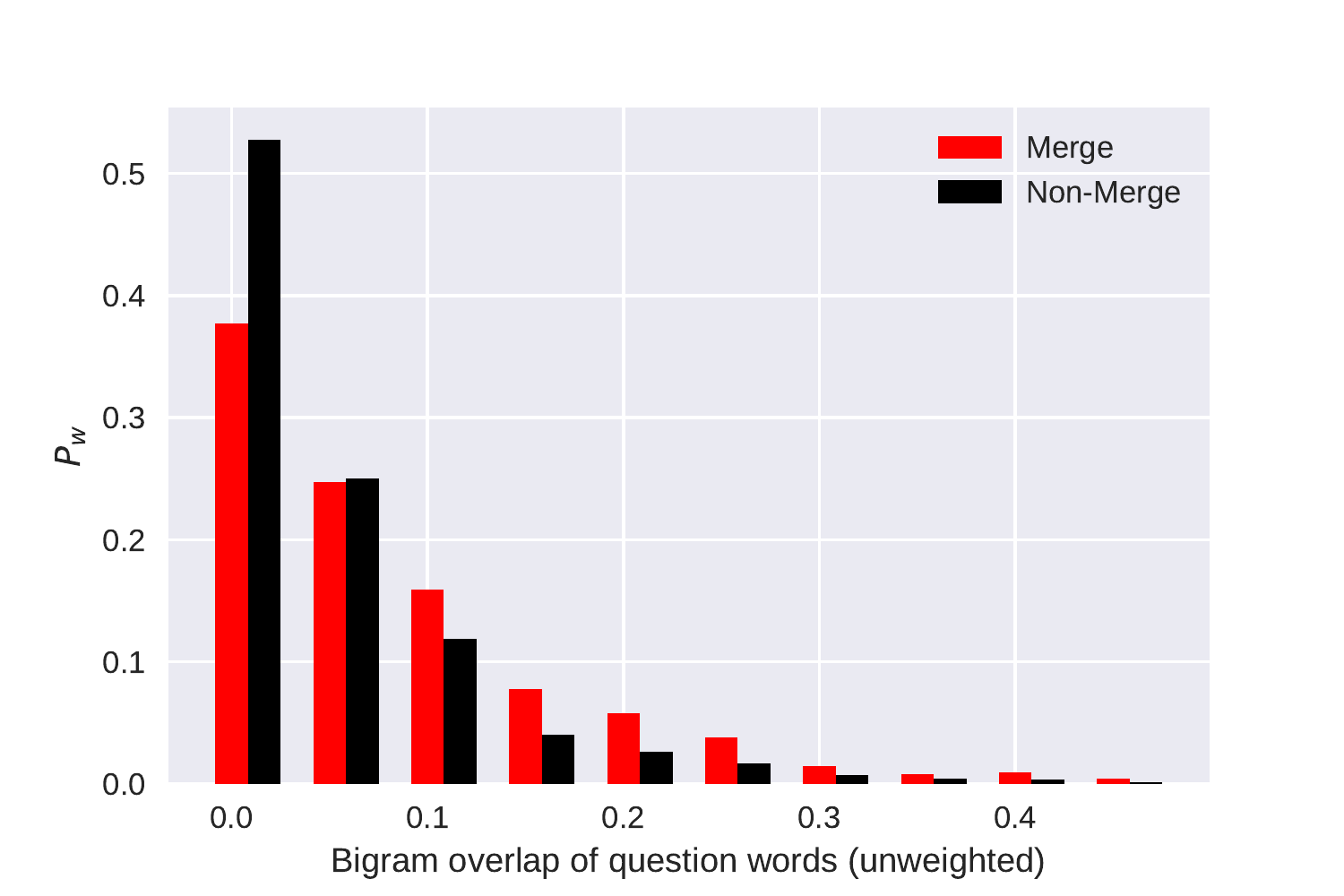} \label{fig:label1}}
	\hfill
	\subfloat[Topic name in question text.]{%
		\includegraphics[width=.33\textwidth]{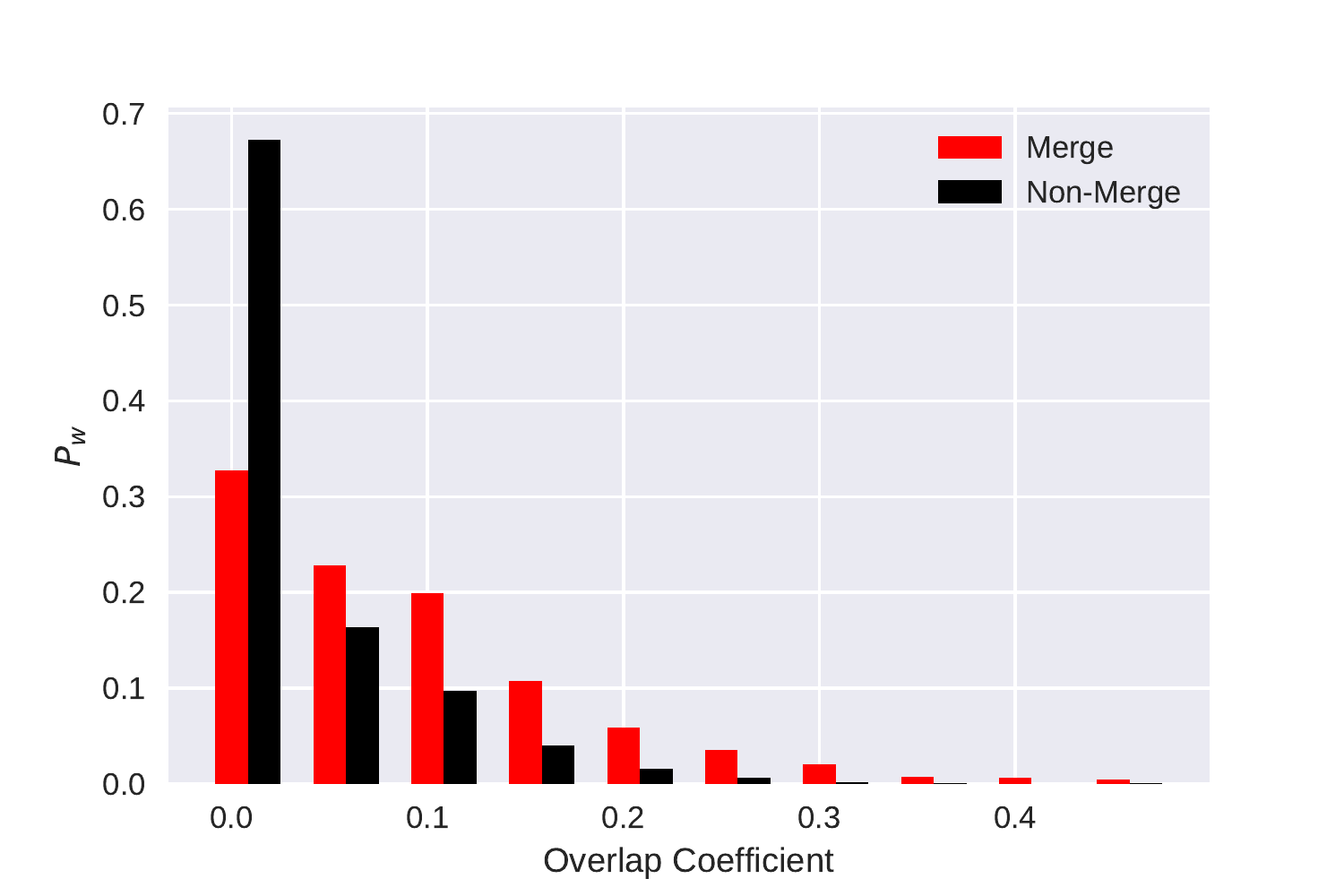}\label{fig:label2}}\hfill
	\subfloat[Co-occurring topic overlap.]{%
		\includegraphics[width=.33\textwidth]{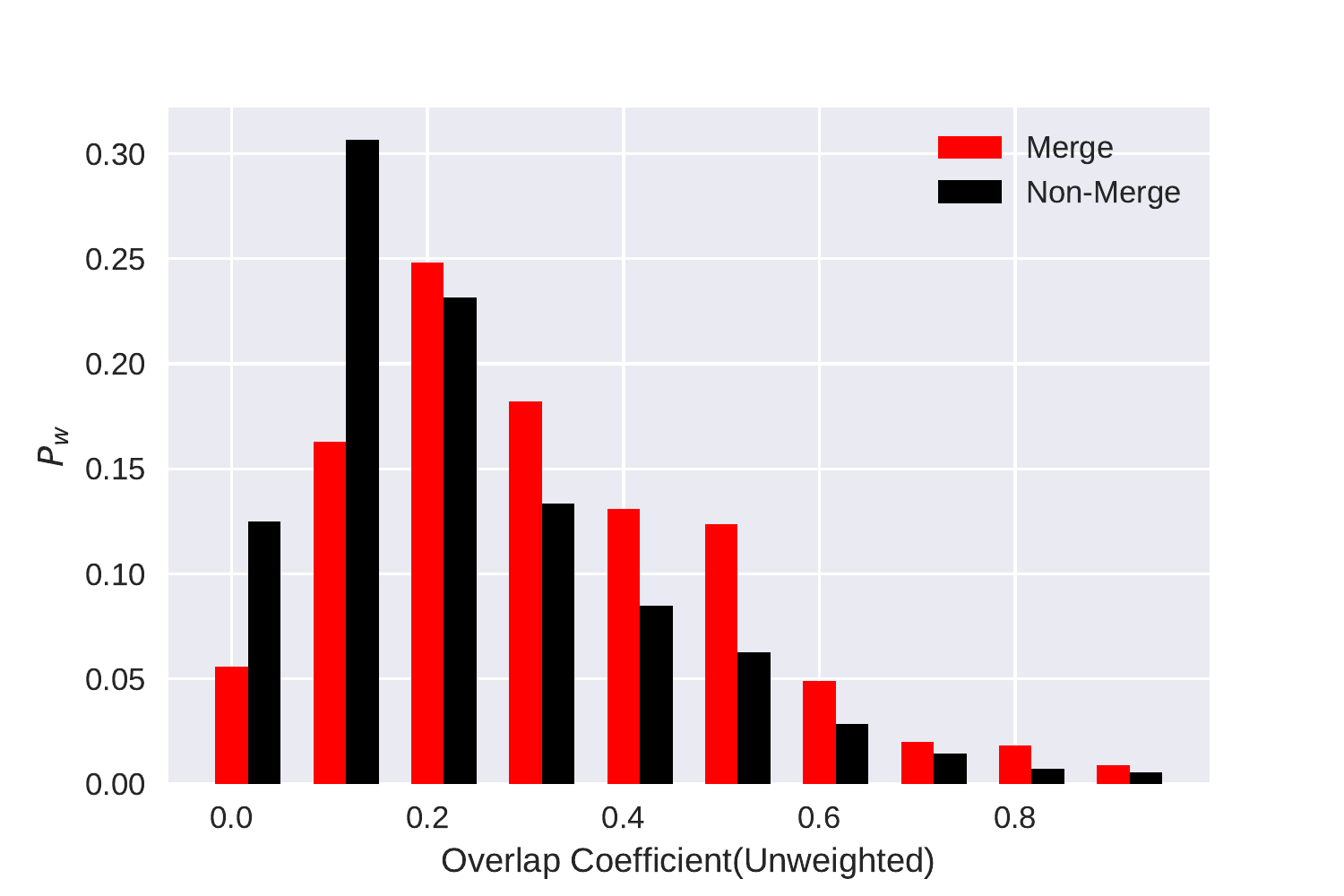}\label{fig:label3}}\hfill
        
	\subfloat[Parent overlap of the co-occurring topics.]{%
		\includegraphics[width=.33\textwidth]{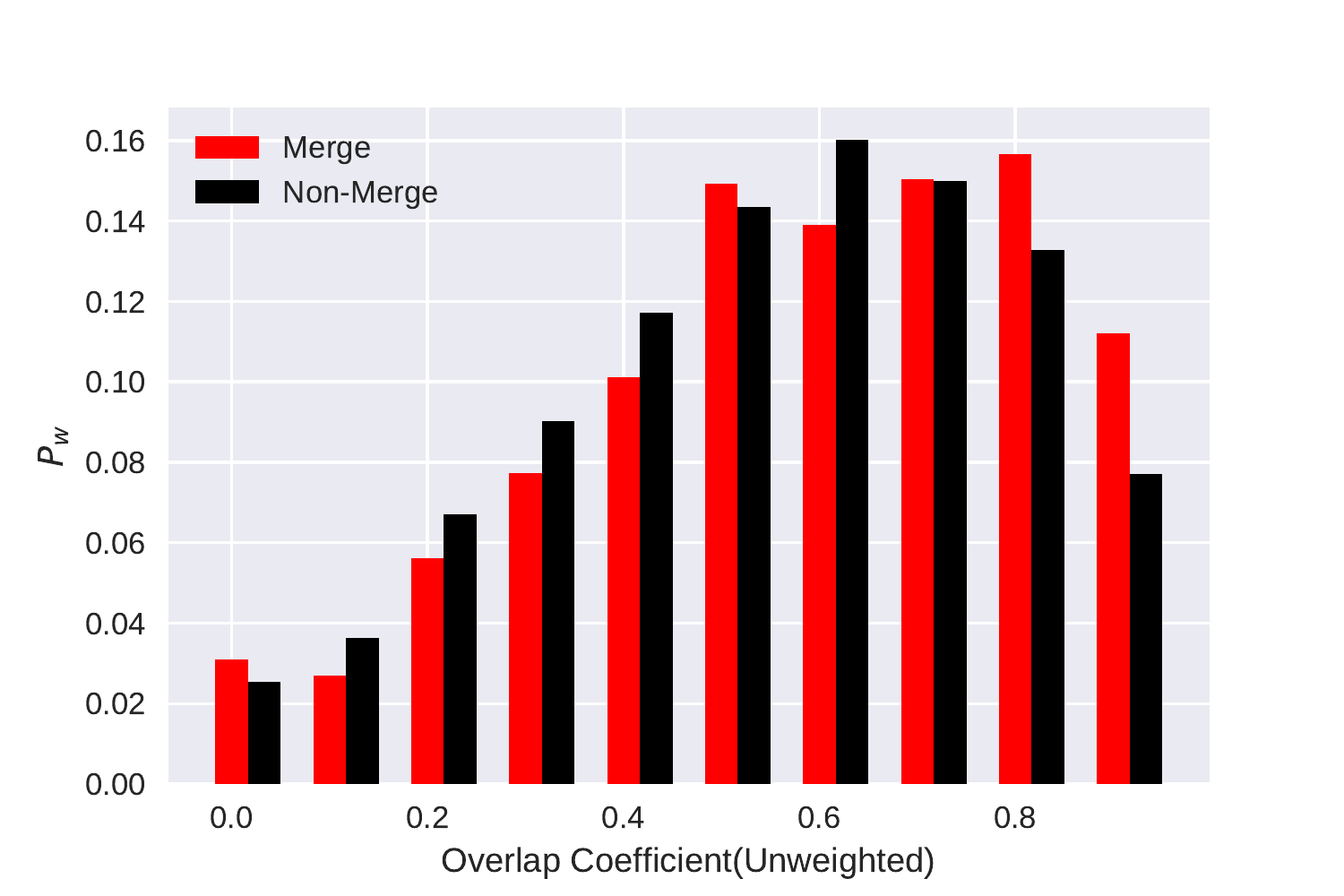}~\label{fig:label4}}\hfill
	\subfloat[Tf-idf similarity value.]{%
		\includegraphics[width=.33\textwidth]{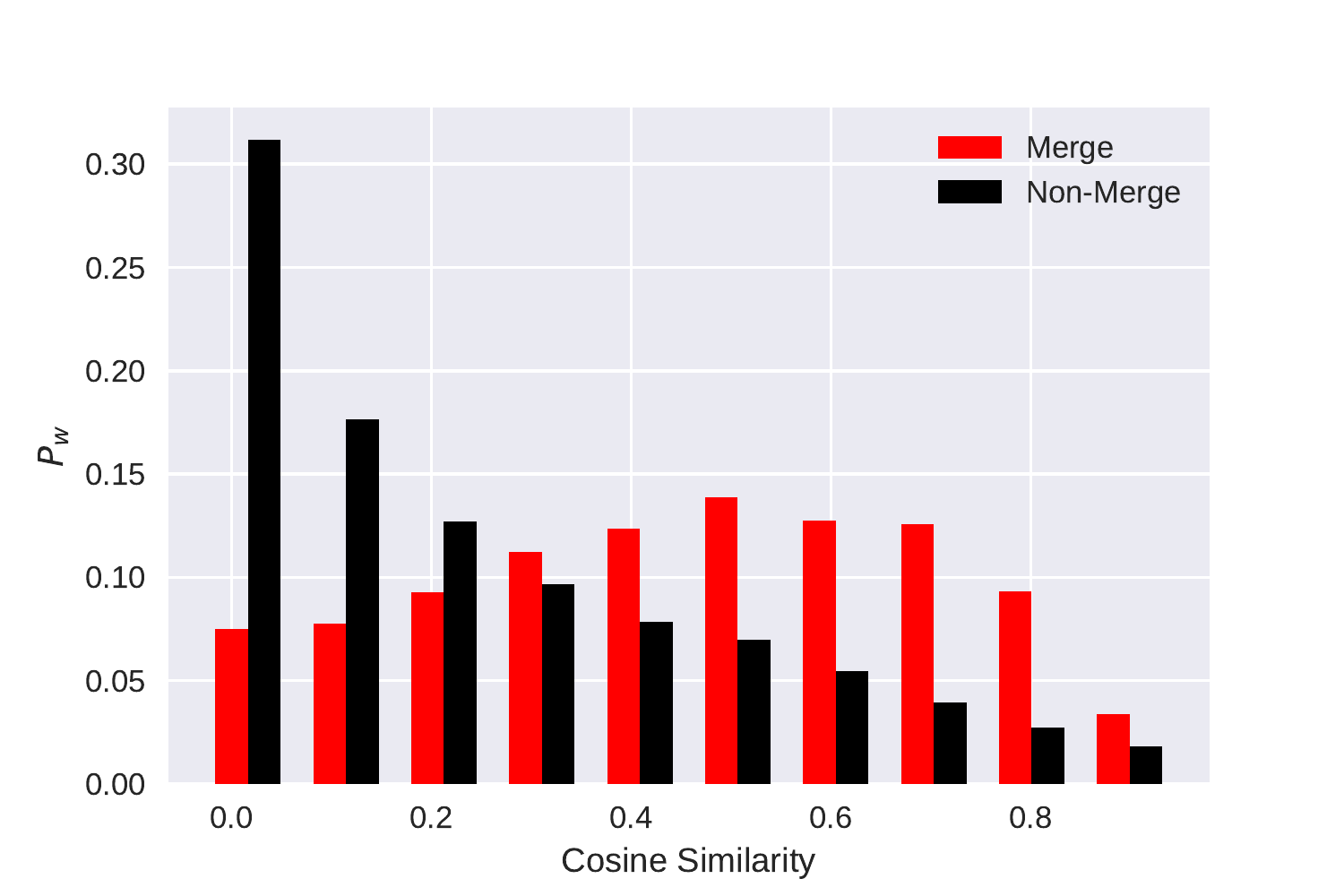}\label{fig:label5}}\hfill
	\subfloat[Normalized Wu-Palmer similarity between question words corresponding to topic pairs.]{%
		\includegraphics[width=.33\textwidth]{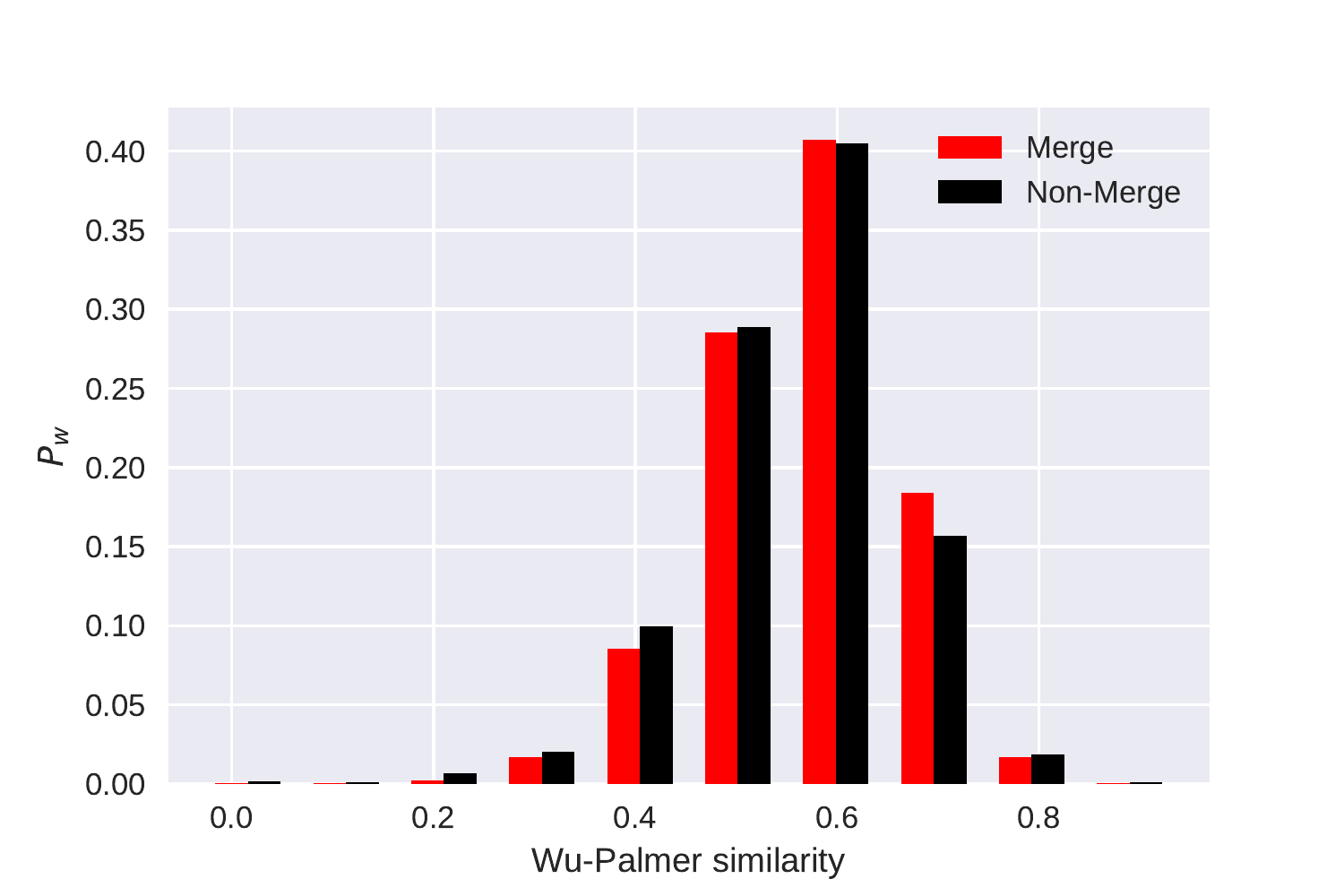}\label{fig:label6}}\hfill  
	
    \subfloat[Average minimum path length of five least frequently co-occurring topic.]{%
		\includegraphics[width=.33\textwidth]{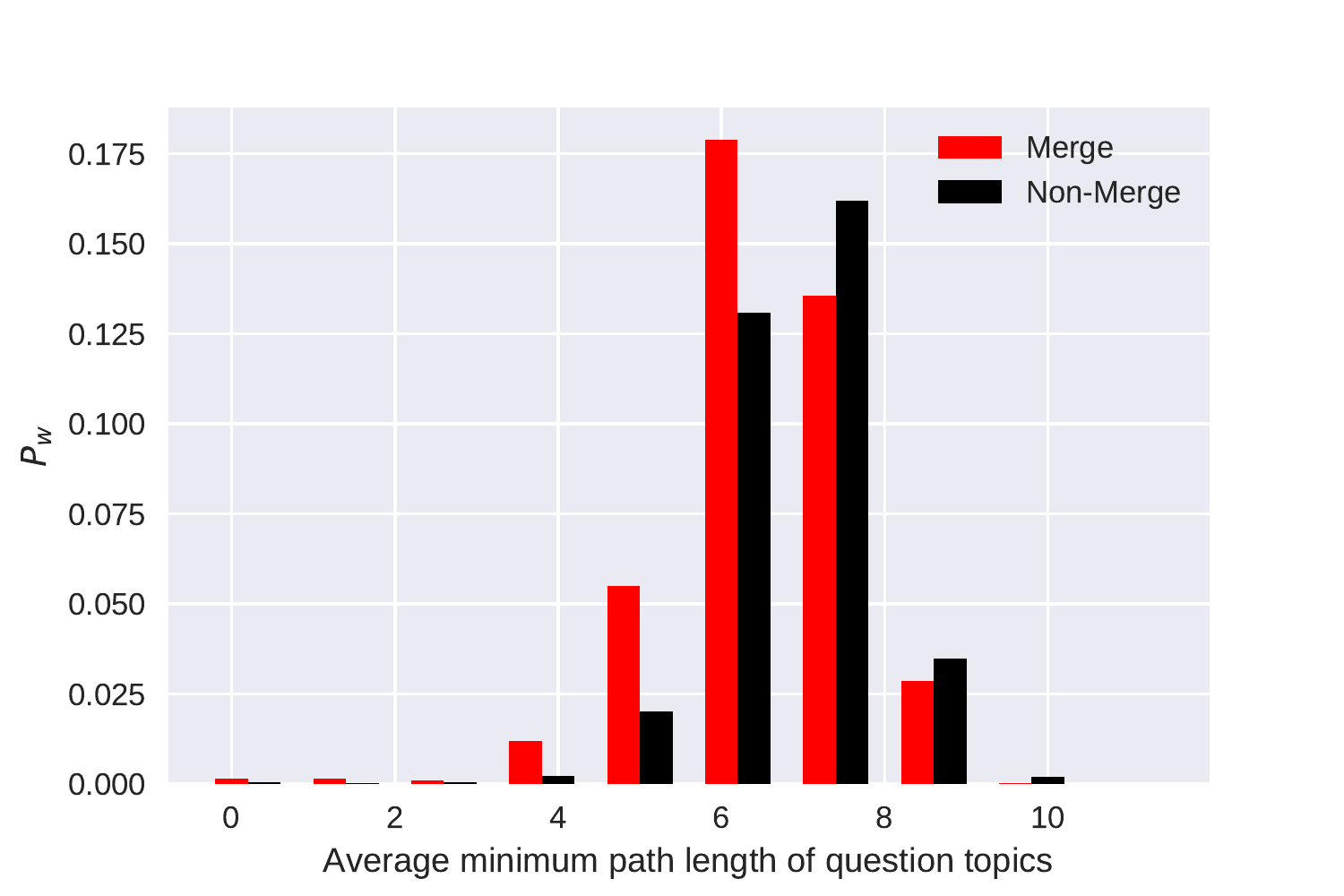}\label{fig:label7}}\hfill
	\subfloat[Adamic/Adar score for the co-occurring topics.]{%
		\includegraphics[width=.33\textwidth]{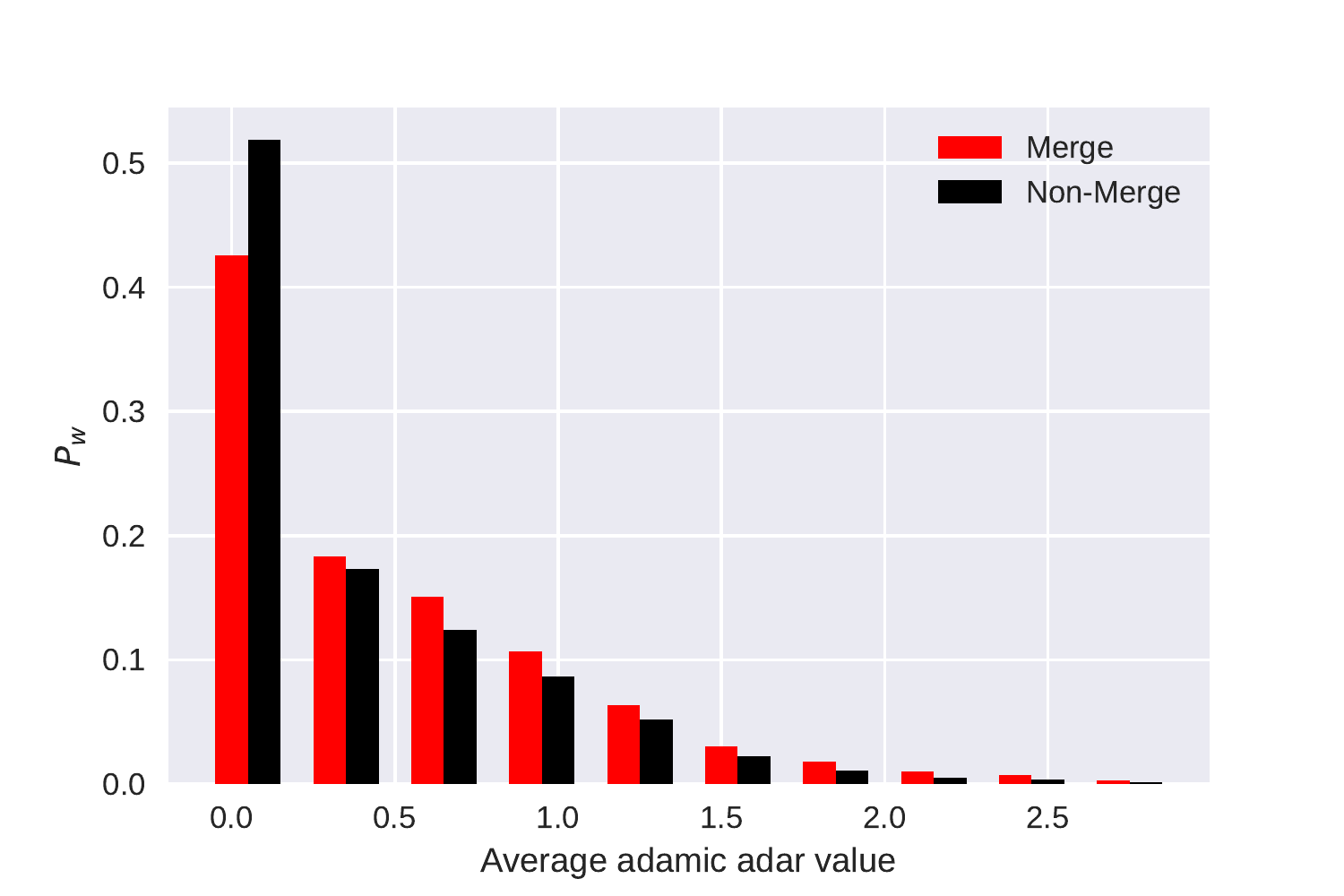}\label{fig:label8}}\hfill
	\subfloat[Lin similarity of top co-occurring question topics.]{%
		\includegraphics[width=.33\textwidth]{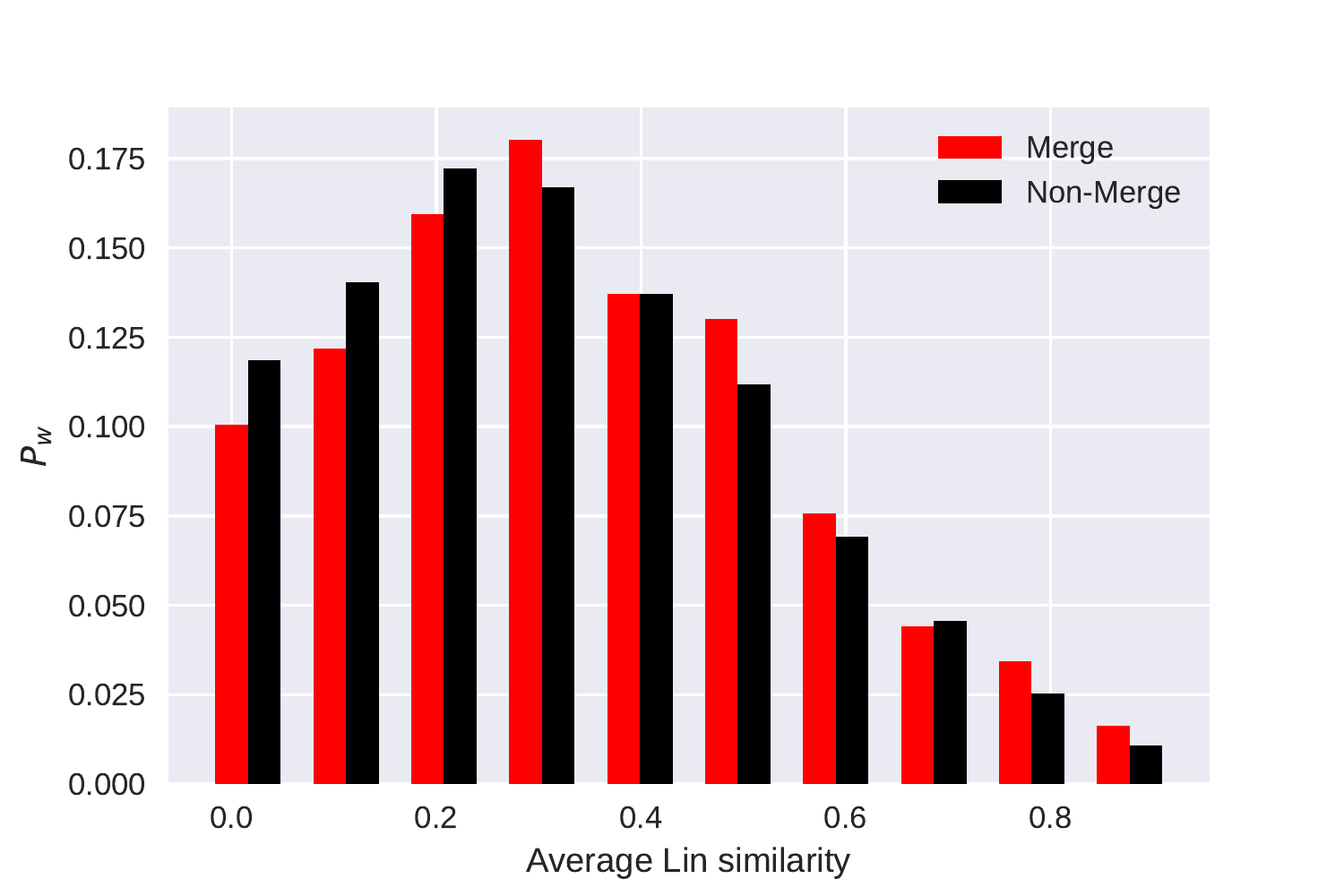}\label{fig:label9}}\hfill
	
    \subfloat[Word2Vec similarity.]{%
		\includegraphics[width=.33\textwidth]{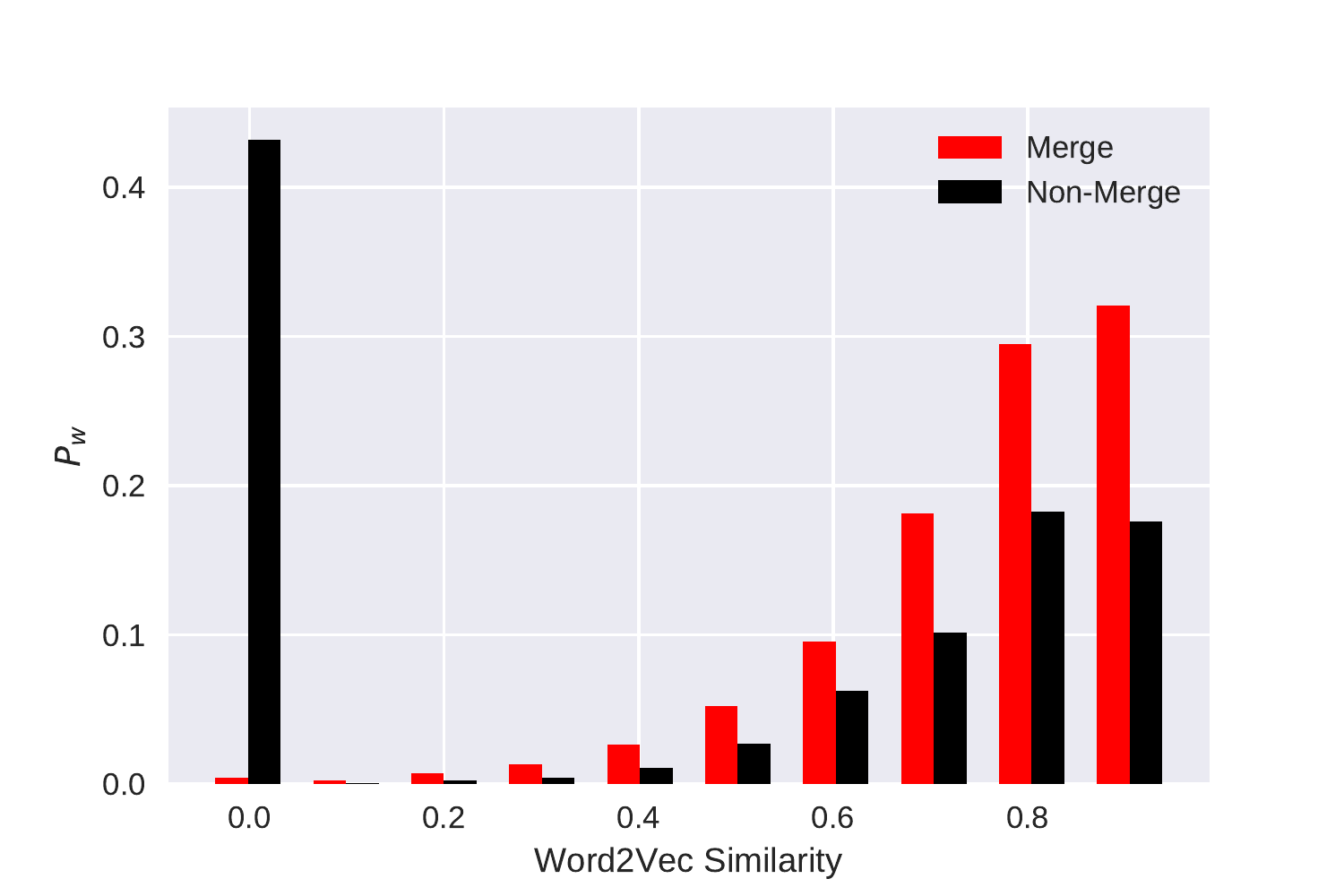}\label{fig:label10}}\hfill
	\subfloat[Doc2Vec similarity.]{%
		\includegraphics[width=.33\textwidth]{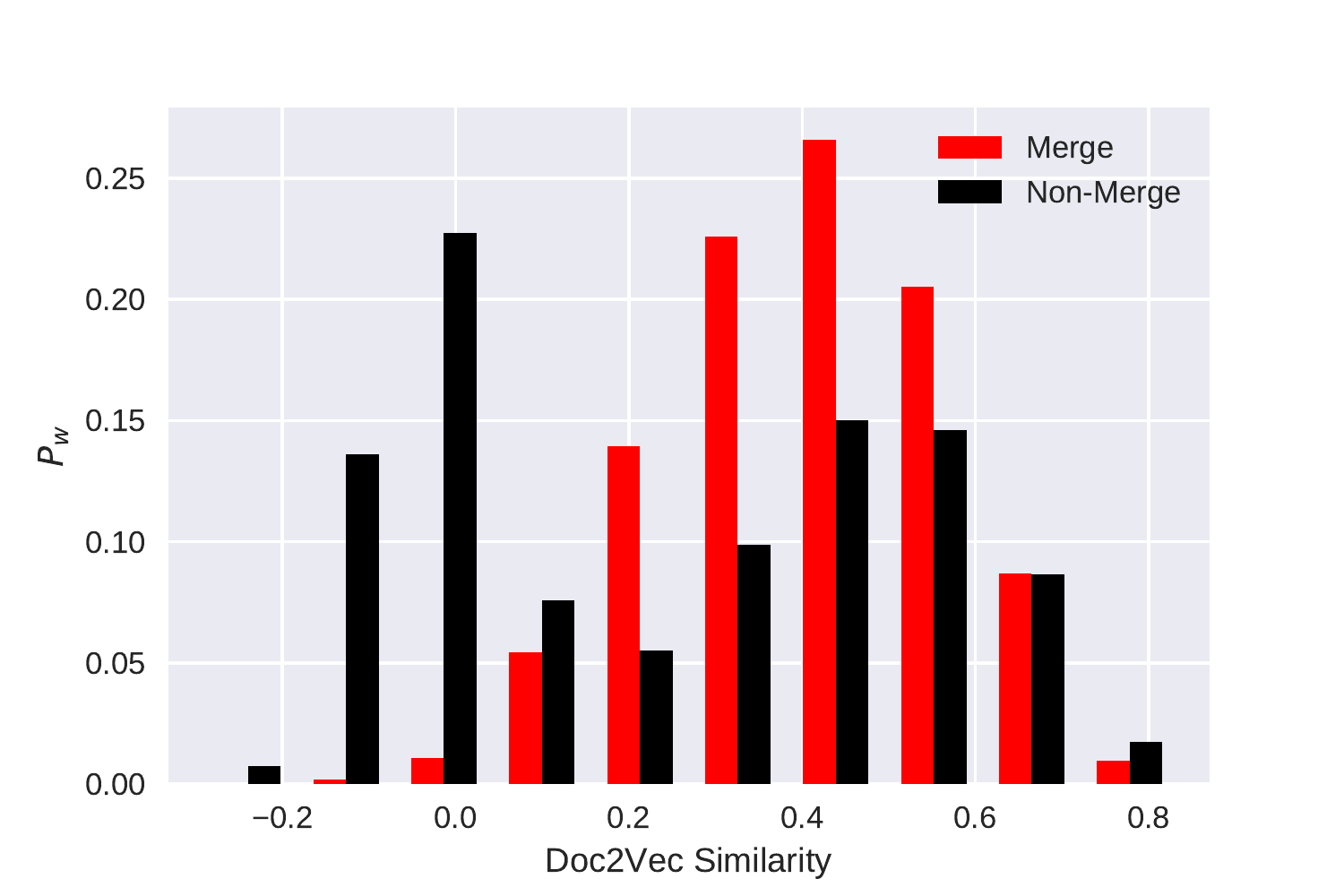}\label{fig:label11}}\hfill
	\subfloat[Part of Speech similarity.]{%
		\includegraphics[width=.33\textwidth]{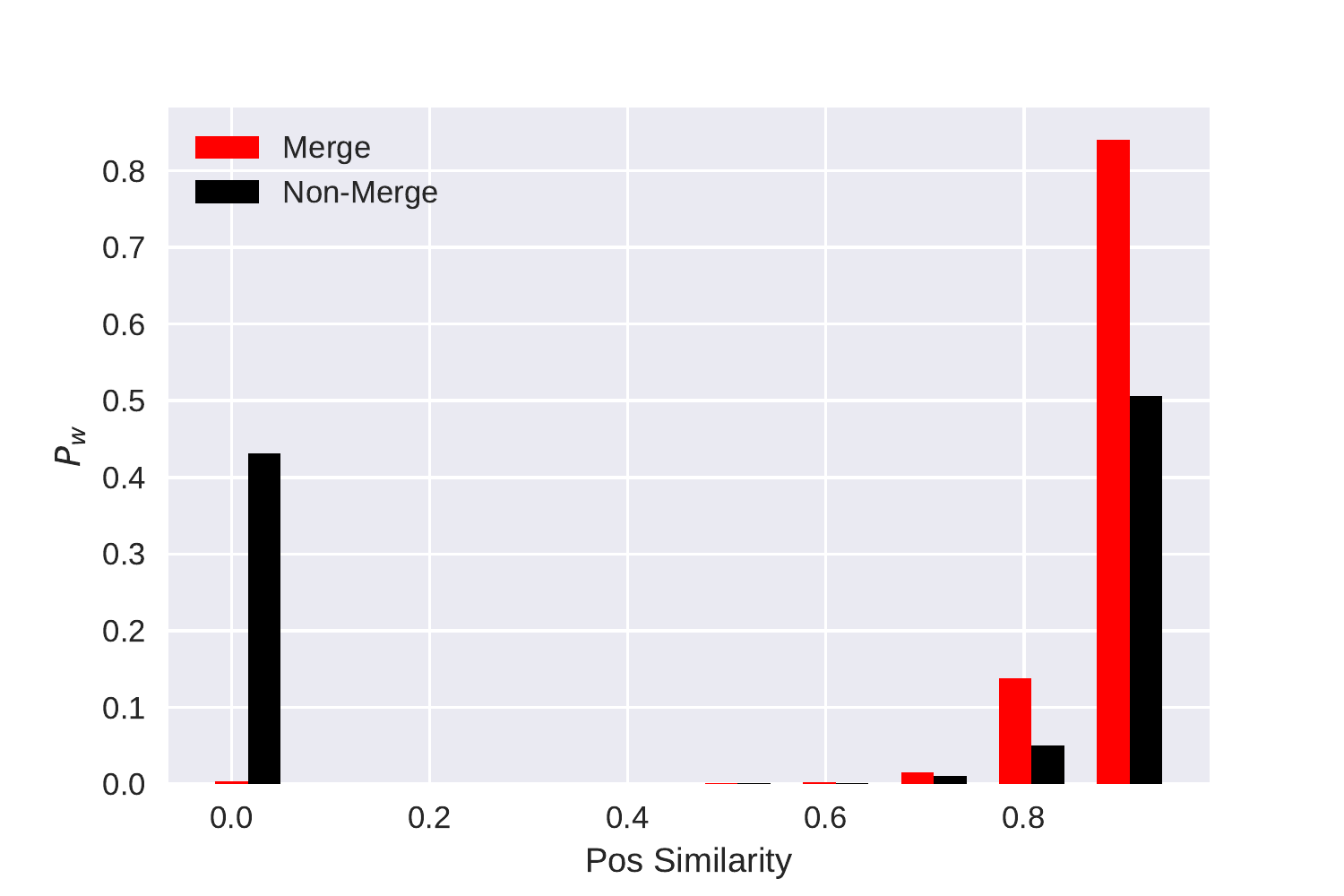}\label{fig:label12}}\hfill
	\caption{Distribution of various properties of the merge and non-merge pairs. The Y-axis represents the fraction of topics ($P_w$) and the X-axis represents the different properties of the topic pairs.}
\end{figure*}

In the weighted version of the overlap, instead of taking each common element once, we consider the minimum frequency of the common element in both the sets. The weighted version of the overlap coefficient is given in Equation~\ref{equ:equation2}
\begin{equation}
\sum_{K\epsilon {X \cap Y}}{\frac{min{\left(K\_freq(X), K\_freq(Y)\right)}}{min \left (\mid X\mid , \mid Y \mid \right)}}
\label{equ:equation2}
\end{equation}

where $K$ represents the set of question text n-gram that are common in the two topics $X$ and $Y$. $K\_freq(X)$ computes the frequency of occurrence of $K$ in Topic $X$ and $\mid X \mid$ represents the number of n-gram elements formed using the questions in Topic $X$.
Figure~\ref{fig:label1} shows the proportion of topics with bigram overlap (unweighted) distribution of question words.  We can observe that non-merge topic pairs have lower overlap coefficient as compared to merge topic pairs.

\noindent\textit{\textbf{Topic name in question text}}: If two topics are describing the same concept, then there is a high probability that this topic name will be present in the question text of the other topic. Based on this hypothesis, we extracted 1, 2, 3 and 4-grams of the topic name and the question words of the other topic. Figure~\ref{fig:label2} shows the unigram distribution of the word overlap of one topic's (topic 1) name and the words present in the other topic's (topic 2) questions. From the figure, we can clearly observe that this hypothesis is true for the merge pairs.

\noindent\textit{\textbf{tf-idf similarity}}: Question similarity between topic pairs is also a factor influencing topics to merge. To obtain question similarity between the topics, we first generate the tf-idf vectors for the questions for both the topics. We then calculate the cosine similarity between them. Figure~\ref{fig:label5} shows the distribution. From the figure, it is evident that non-merge topics have lower similarity values as compared to the merge pairs.

\noindent\textit{\textbf{Co-occurring topic overlap}}: Each topic is associated with a set of questions and each such question is associated with other topics. These topics are called the co-occurring topics. For example, a question about sports may be tagged with `sports', `athletes', `exercise', `practise'. Thus, the topics `sports',`athletes' and `practise' are the co-occurring topics of `exercise'.
For each topic pair, we calculate the weighted and unweighted overlap coefficient of the co-occurring topics. Figure~\ref{fig:label3} shows the distribution of the unweighted overlap coefficients of the co-occurring topics. We can observe that the non-merge topics tend to have lower overlap coefficient as compared to the merge topics. The results for the weighted overlap are very similar and therefore not shown.

\noindent\textit{\textbf{Word2Vec similarity}}: We use Word2Vec~\cite{mikolov2013efficient} to generate word embeddings for the words in the question text corresponding to a topic. The experiments were performed using gensim~\cite{rehurek_lrec}. We set minimum word count to 5 and the size of feature vector to 150. We then generate a word vector for a document (i.e., all questions associated to a topic) by averaging all the vectors of the question words in that topic. After this, we perform cosine similarity to determine the semantic similarity of the two topics. It is evident from Figure~\ref{fig:label10} that the merge pairs have a higher similarity between their word vectors.

\noindent\textit{\textbf{Doc2Vec similarity}}: We use Doc2Vec~\cite{le2014distributed} to generate vectors for each document. A document here consists of all questions associated to a particular topic. The experiments were performed using gensim~\cite{rehurek_lrec}. We set minimum word count to 5 and the size of feature vector to 150. The cosine similarity is calculated between the documents corresponding to a topic pair. Once again, the merge pairs seem to have higher overall similarity between their corresponding document vectors (see Figure~\ref{fig:label11}). 

\noindent\textit{\textbf{Part-of-speech}}: We use the Stanford POS tagger~\cite{manning2014stanford} to find the part of speech tag for each question word associated to a topic. For each topic we build a POS tag vector containing the count of the different POS tags in the associated questions. We then compute the similarity between two topics as the cosine similarity between the POS tag vectors of the two topics. This similarity is much higher for the merge pairs (see Figure~\ref{fig:label12}).

\subsubsection{Quora topic ontology}
Quora topic ontology provides parent-child relationships between topics. We derived various features that make use of the ontology to show the contrast between the merge and the non-merge pairs.

\noindent\textit{\textbf{Co-occurring topics' parent and child overlap}}:
Apart from overlap between the co-occurring topics of the topic pair, we also calculate the parent and child topic overlap (weighted and unweighted) of the co-occurring topics. Figure~\ref{fig:label4} shows the distribution of the co-occurring topics' parent-child overlap (weighted) and it is evident from the plot that the merge pairs tend to have higher overlap as compared to the non-merge topic pairs.

\noindent\textit{\textbf{Path length between topics}}: In our merge dataset, only 17.4\% of the pairs are part of the topic
hierarchy. Rest of the pairs have at least one topic which is not part
of the ontology graph. Therefore, calculating the path length between
the topics directly is not very useful. Instead, we use the average
minimum path length between the co-occurring topics. We first find the
five most frequent co-occurring topics of each pair and then we
calculate the average of the minimum path length between all pairs of
these co-occurring topics. The distribution of the five least common topics in the
merge and non merge pairs is presented in Figure~\ref{fig:label7}. We
can observe that merge pairs tend to have lower average distance as
compared to non-merge pairs. 

\noindent\textit{\textbf{Adamic/Adar similarity between co-occurring topics}}: Adamic / Adar similarity~\cite{adamic2003friends} is defined as inverted sum of degrees of the common neighbors for two topics in the Quora topic ontology graph.
\begin{equation}
\sum_{u\epsilon {N(x) \cap N(y)}}{\frac{1}{\log_{}{(\mid{N(u)}\mid)}}}
\end{equation}
where $N(u)$ is the set of neighbors of vertex $u$. We compute the
Adamic/Adar similarity value between the co-occurring topic
pairs. Figure~\ref{fig:label8} shows the distribution of the average
Adamic/Adar value for the ordered pairs of five most frequently
co-occurring topics for each candidate topic pair. As usual, the merge
pairs have higher Adamic/Adar values.

\noindent\textit{\textbf{Similarity measures on the ontology}}: We also calculate semantic similarity between topics to distinguish
merge and non-merge topic pairs. Semantic similarity of a set of
topics is a metric in which the idea of distance between the two
topics estimates the strength of the semantic relationship between the
topics. We use the Quora topic ontology to find the similarity between
two topics. Specifically, we use the Lin
similarity~\cite{lin1998information}, Resnik
similarity~\cite{resnik1999semantic}, Wu-Palmer
similarity~\cite{wu1994verbs} and JCN
similarity~\cite{jiang1997semantic} for this
task. Figure~\ref{fig:label9} shows the distribution of average Lin
similarity of five most frequently co-occurring topics for each merge
and non-merge pair. The result shows that the merge pairs are
semantically closer than the non-merge pairs.

\subsubsection{External similarity features}

\noindent\textit{\textbf{Wordnet similarity}}:
We use several
similarity measures defined over the Wordnet to calculate the
similarity. For each topic pair, we calculate the similarity between the question words corresponding to the two
topics. Figure~\ref{fig:label6} shows the similarity distribution for normalized Wu-Palmer. We
can see that non-merge pairs have lower similarity as compared to
merge pairs showing that the merge pairs have questions that are
indeed semantically more close.

\section{Prediction framework}

The topics in Quora allow the community to provide good quality responses to the questions as well as to organize them properly. With the number of topics in Quora increasing continuously, the number of duplicate topics are increasing as well as observed in Figure~\ref{fig:merge_unmerge_timeline}. A Quora user on average takes 936 days to find a duplicate topic and merge it with the original. One of the main reasons for this delay in merging is the huge number of topics in the Quora ecosystem. It would require a lot of effort to manually inspect each pair of topics and determine, if they should merge or not. We propose to build a system that would help the user in this task. As the number of topic pairs possible could be millions, we cannot directly use a supervised classification task. Instead, we propose a two-step approach combining the \textit{anomaly detection} and the \textit{supervised classification framework} to automatically predict whether two given topics would merge or not in future using the features we discussed in the above section. The positive instances for our approach are those topic pairs which should merge as they represent the same concept. The negative class, on the other hand, are the set of topic pairs which do not represent the same concept. The entire flow of the approach is illustrated in Figure~\ref{fig:classification_flowchart} and described below.

\subsection{Merge prediction}
\noindent\textbf{Positive training and test instances for the supervised classification task}: We divide the merge instances into 7:3 ratio chronologically. This means that 70\% (1981) of the merge cases that occur before \textit{Feb 21, 2016} are considered as positive training instances for the supervised classification and the rest 30\% (849) of the merges which occur after this time point are considered as positive test instances.

\noindent\textbf{Negative training instances for the supervised classification task}: For the negative training instances of the supervised classification we consider the unmerge pairs and the neighborhood pairs (i.e., pairs in the non-merge class).

\noindent\textbf{Negative test instances for the supervised classification task}: In the application scenario, we need to predict the pair of merged
topics from among all the possible topic pairs. For the negative test set,
therefore, we want to use all possible pairs of topics but with a
set of appropriate filters. We describe these filters below.

\noindent{\em Filters:} As a first filter, we remove all those topics which had less than 20 questions in them. This brought down the number of topics to 45,743 (i.e., close to 2 billion topic pairs). Next, we apply the same set of filters that we applied to the merge topic pairs (removal of abbreviations and high Jaro-Winkler similarity topic pairs). This reduced the number of pairs to one billion. Further, we calculate the overlap coefficient of the co-occurring topics of each pair. We use a threshold of 0.25 to obtain around 20 million negative test pairs. We find that this filter allows 89\% of the positive merge pairs. Note that we use all these filters to make the negative class more non-trivial. Since it was computationally almost impossible to
generate all the possible features for the 20 million negative
instances, we randomly select one million negative instances for the testing purpose.

\begin{figure}[tbp]
	\centering
	\includegraphics[width=.50\textwidth]{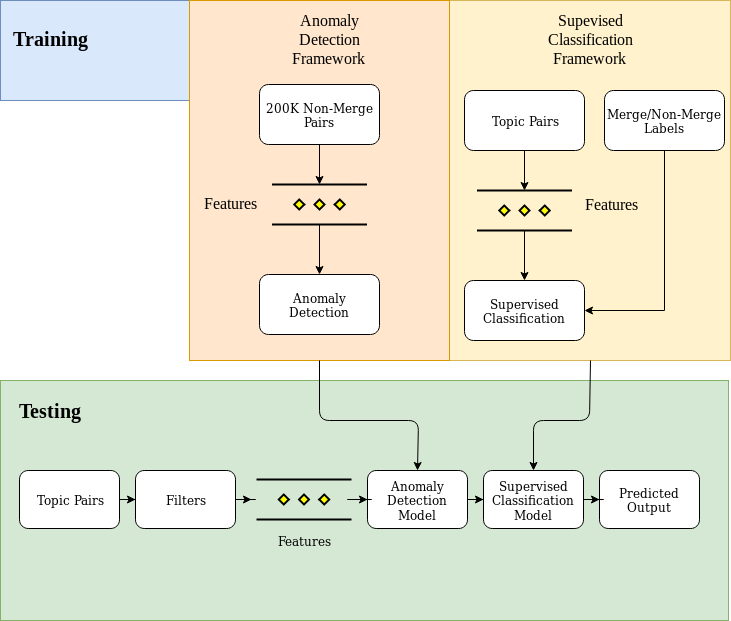} 
	\caption{Flowchart of the two step classification approach. The anomaly detection framework acts as a filter which removes substantial negative test instances while retaining a good recall score on the positive test set. These are then passed to the supervised classifier for prediction.}
    \label{fig:classification_flowchart}
\end{figure}
\noindent\textbf{Reducing the test instances for the supervised classification task}: It is here that we use anomaly detection algorithms to obtain a reduced set of test instances that we can feed to the supervised classifier. In particular we generate 200K non-merge pairs (fully distinct from those used as negative examples for training the supervised classifier) to train the anomaly detection algorithms. Using this, we identify those test instances from the 1M (negative test) + 849 (positive test) topic pairs that are `outliers', i.e., closely resemble the properties of the merge class and are therefore, strong candidates of merging. We retain these instances in the final test set for the supervised classifier. The key objective is to maintain a high recall while allowing for some false positives.

\noindent\textbf{Models used for anomaly detection}: We use standard anomaly detection methods such as One-Class SVM (OCSVM), Isolation Forest (IF), Local Outlier Factor (LOF), Elliptic Envelope (EE) for our experiments. We set contamination to 0.2 in all the experiments. For OCSVM, we set the kernel as `rbf'. For IF we set the number of estimators as 100. For LOF, we set the number of neighbors as 20.

\noindent\textbf{Models used for supervised classification}: The classification methods used include Random Forest (RF), Naive Bayes (NB), Support Vector Machine (SVM), Logistic Regression (LR), Decision Tree (DT), XGBoost (XGB). Since our training data is imbalanced we use a cost-sensitive classification approach. In case of Random Forest (RF) we set the number of tree as 10. For SVM, we user Linear Kernal and C as 1.0. For LR, we use `l2' penalty with C as 1.0 and set `liblinear' as the optimization algorithm. For XGB, we set the learning rate as 0.1, booster as gbtree and use 100 boosted trees to fit.  

\noindent{\bf The overall two-step method}: As discussed above we first use an anomaly detection algorithm followed by a supervised classification approach. The anomaly detection framework acts as a filter which removes substantial negative test instances while retaining a good recall score on the positive test set. These are then passed to the supervised classifier for prediction. The whole process is explained in Figure~\ref{fig:classification_flowchart}.

\noindent{\bf Competing baselines}: Examples of automatic merge prediction of linguistic entities is rare in the literature. We use two baselines to compare the performance of our model. We found that Maity et al.~\cite{MaitySM16} has resemblance with our work in this respect and therefore, used it as the first baseline. We adapt the feature set which are appropriate for the topic merge and compare the performance. 
For the second baseline, we use Universal Sentence Encoder~\cite{cer2018universal} to generate 512 dimensional vector representation of the topics. For each topic, we first concatenate all the questions and pass it to the encoder to generate the 512 dimensional vector representation for a topic. In order to determine if a topic pair is a merge or not, we simply calculate the cosine similarity between their vector representations. In order to distinguish between the merge and non-merge topic pairs, we define a threshold $T$ for the similarity, above which we would call a topic pair as merge and below which it will be a non-merge. We use the training dataset to set the value of $T$ to be the value of cosine similarity which maximizes the F-score. We found that a threshold value of $0.9$ performs the best.

\begin{table}[!htb]
	\centering
	
	\resizebox{9cm}{!}{%
		\begin{tabular}{|l| l  l l l |} 
			\hline
			Type & Algorithm &Precision & Recall & F-Score \\ [0.5ex] \hline\hline 
			\multirow{2}{*}{Baseline}&Maity et al.~\cite{MaitySM16} &0.192    &   0.167 & 0.179\\
			&Universal Sentence Encoder~\cite{cer2018universal} & 0.552 & 0.290  &  0.380 \\[0.5ex] \hline \hline
			
			\multirow{4}{*}{Outlier Detection}&Local Outlier Factor &0.012&0.814&0.024\\
			&\cellcolor{green}Isolation Forest &\cellcolor{green}\textbf{0.041}&\cellcolor{green}\textbf{0.821}&\cellcolor{green}\textbf{0.078}\\
			&One Class Support Vector Machine  &0.002&0.498&0.004\\
			&Elliptic Envelope  &0.013&0.917&0.026\\ [0.5ex] \hline \hline 
			
			\multirow{6}{*}{Supervised Classification}&Random Forest &0.328&0.491&0.393\\
			&Naive Bayes &0.374&0.075&0.125\\
			&Support Vector Machine &0.069&0.667&0.125\\
			&\cellcolor{green}Logistic Regression  &\cellcolor{green}\textbf{0.325}&\cellcolor{green}\textbf{0.585}&\cellcolor{green}\textbf{0.418}\\
			&Decision Tree  &0.003&0.901&0.005\\ 
			&XGBoost &0.066&0.522&0.117\\ [0.5ex] \hline \hline

			\multirow{5}{*}{Our 2-step method}&Elliptic Envelope + Logistic Regression &0.773&0.527&0.627 \\ 
			&Isolation Forest + Random Forest &0.706&0.624&0.662 \\
			&\cellcolor{green}Isolation Forest + Logistic Regression &\cellcolor{green}\textbf{0.866}&\cellcolor{green}\textbf{0.603}&\cellcolor{green}\textbf{0.711} \\
			& Isolation Forest + Support Vector Machine &0.833&0.491&0.618 \\
			& Local Outlier Factor + Logistic Regression &0.612&0.597&0.604 \\ \hline

		\end{tabular}%
	}
	\caption{Results of topic merge prediction.}
	\label{tab:topic_merge}
\end{table}

\begin{table*}[ht]
	\parbox{.45\linewidth}{
		\centering
		\footnotesize 
		\begin{tabular}{p{37mm} p{9mm} p{6mm} p{9mm}} 
			\hline
			Feature Model & Precision & Recall &F-Score \\ [0.5ex] 
			\hline\hline
			All                          &    0.866 &  0.603  & 0.711 \\
			\cellcolor{green}Question Content + Ontology  & \cellcolor{green}\textbf{0.855} & \cellcolor{green}\textbf{0.592} & \cellcolor{green}\textbf{0.700} \\
			Question Content + External  &   0.937 & 0.548 & 0.691 \\
			Ontology + External          & 0.648 & 0.056 & 0.103 \\
			Question Content & 0.921 & 0.533 & 0.675\\
			[1ex] 
			\hline
		\end{tabular}
		\caption{Results of ablation experiments.}
		\label{tab:abalation}
	}
	\hfill
	\parbox{.45\linewidth}{
		\centering
		\footnotesize 
		\begin{tabular}{lll}
			Rank & Feature Name \\ \hline
			1 & Topic name in question text (unweighted)\\
			2 & Question topic overlap\\
			3 & Question topic parent child overlap\\
			4 & Question words overlap top 20\%\\
			5 & Topic name in question text (weighted)\\
			6 & Word2Vec Similarity\\
			7 & Topic parent child overlap\\
			8 & Tfidf Similarity \\
			9 & Bottom topics average minimum path length\\
			10 & Top 5 topics Adamic/Adar measure\\
			\hline
		\end{tabular}
		\caption{Importance of features.}
		\label{tab:feat_importance}
	}
\end{table*}

We have several interesting observations. First, we find that the precision of all the anomaly detection algorithms are very low. Second, the classification methods seems to perform better than the anomaly detection framework especially in precision and F-score. Third, we find that while the universal sentence encoder seems to be a stronger baseline than Maity et al.~\cite{MaitySM16}, the two-step approach performs the best and outperforms both the baselines by a huge margin. The best performance is obtained when we use isolation forest for anomaly detection followed by logistic regression for the classification. Our system could be used to improve the existing workflow at Quora by suggesting possible topics which represent the same concept. This would allow the community to detect and merge topics in an early stage of the topic evolution, in turn, enabling timely and appropriate knowledge aggregation.

\noindent\textbf{Ablation experiments}: 
To understand the importance of individual feature types, we perform
ablation experiments by considering different combination of feature
types for the classification. Table~\ref{tab:abalation} presents the results for the test set containing
849 positive instances and 1M negative instances applied on the two-step framework using isolation forest followed by logistic regression (IF+LR). The combination of question content and
the ontology feature types seems to produce results, very close to the best results, obtained by taking all features.

\noindent\textbf{Feature importance}: 
 Table~\ref{tab:feat_importance} shows the ranking (based on recursive feature elimination algorithm ~\cite{guyon2002gene}) of the top 10 features. The most discriminatory feature is the presence of topic name in the question text of the other topic. Also Quora topic ontology features have higher ranks.


\subsection{Direction of topic merge}
When a user decides to merge two topics, he/she has to make a choice about the direction of merge. The direction decides the outcome of the competition between conventions.  When two topics get merged, one of them ceases to exist and would no longer be accessible to the users directly. Let us assume that `Topic A' gets absorbed into `Topic B' to form the new `Topic B', i.e., the winning convention. In the following we discuss the factors instrumental in influencing the direction of topic merge.

\begin{figure*}[!htb]
	\centering
	\subfloat[Number of characters.]{%
		\includegraphics[width=0.24\textwidth]{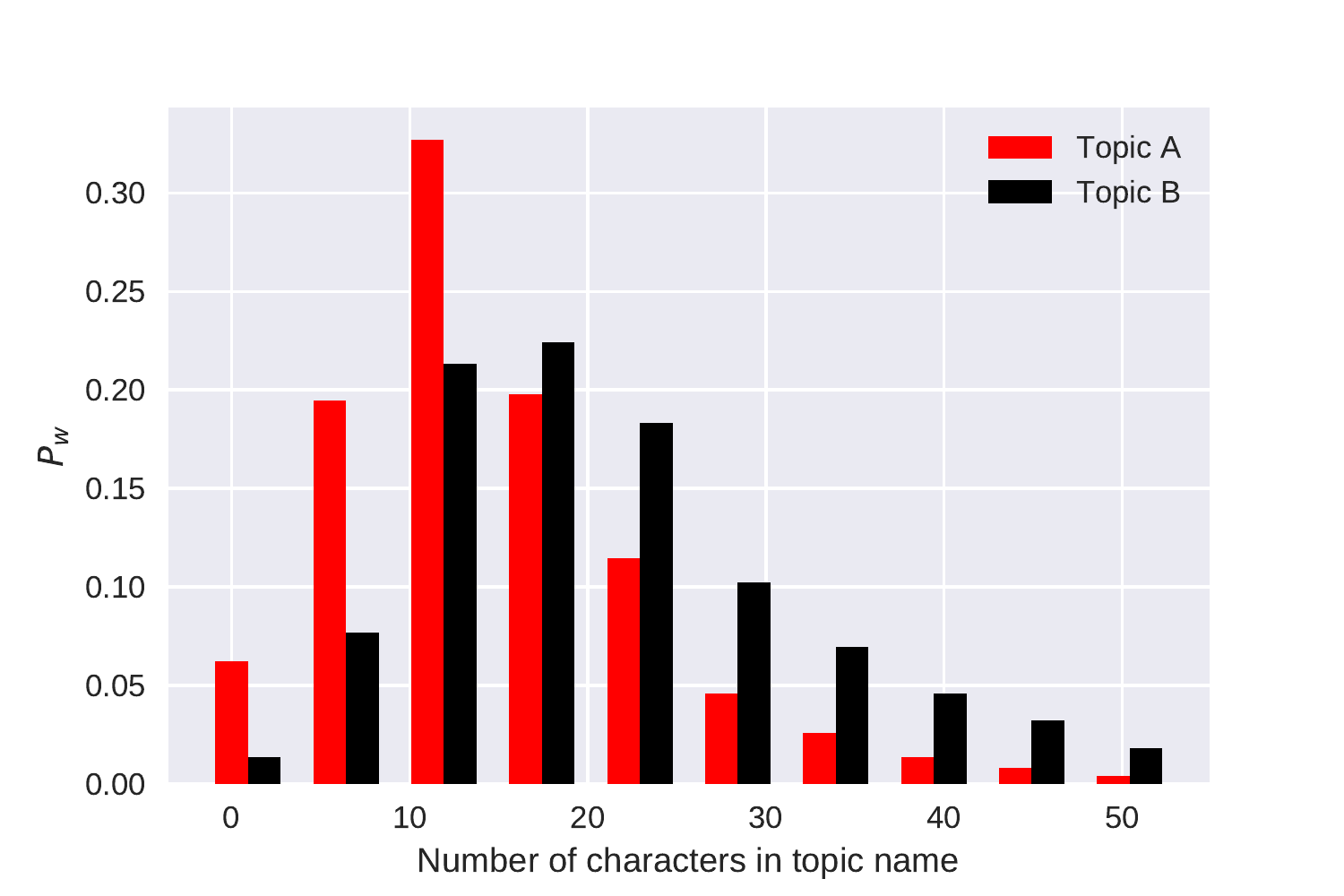} \label{fig:dir_fig1}}
	\subfloat[Number of words.]{%
		\includegraphics[width=0.24\textwidth]{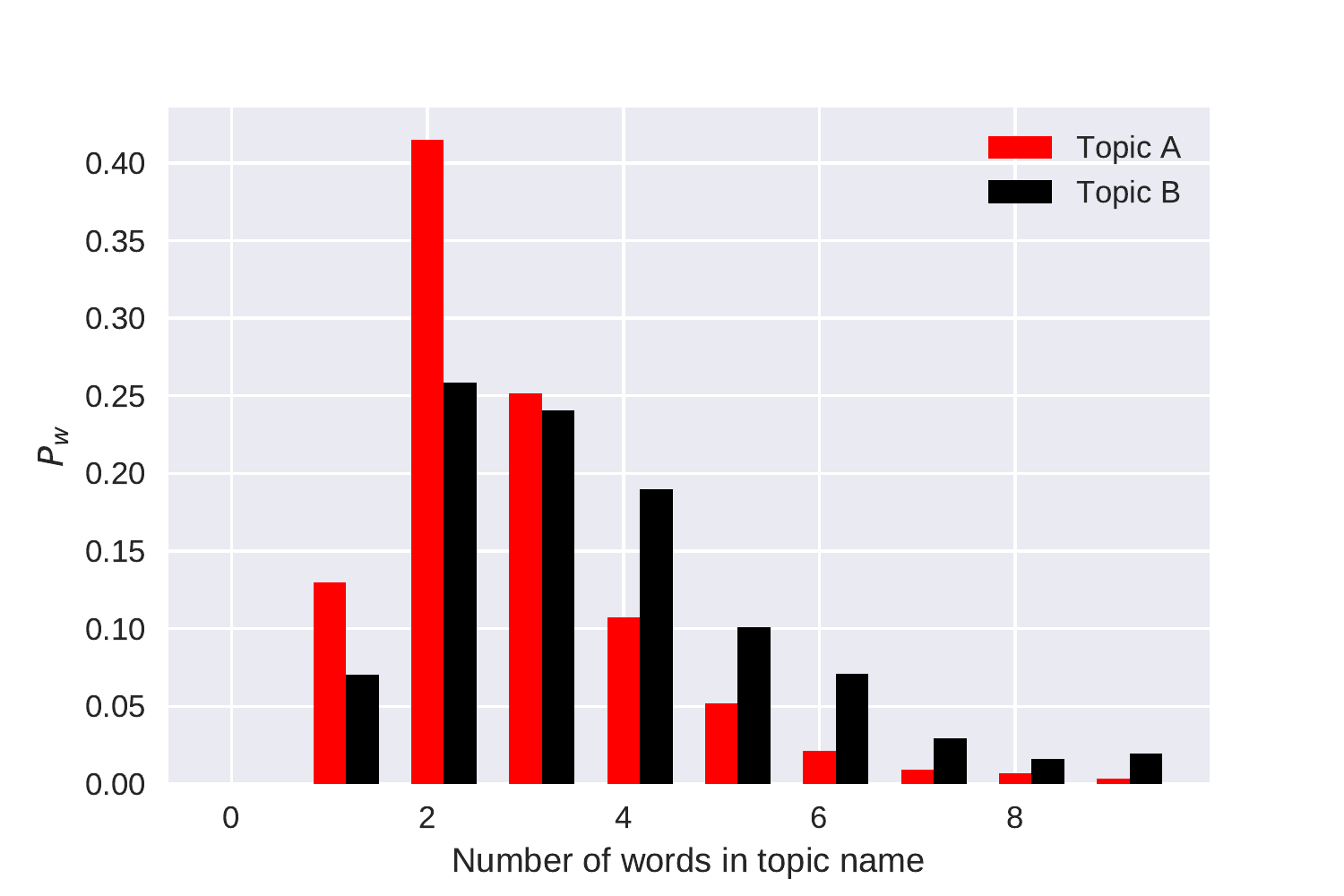}\label{fig:dir_fig2}}
	\subfloat[Number of questions]{%
		\includegraphics[width=0.24\textwidth]{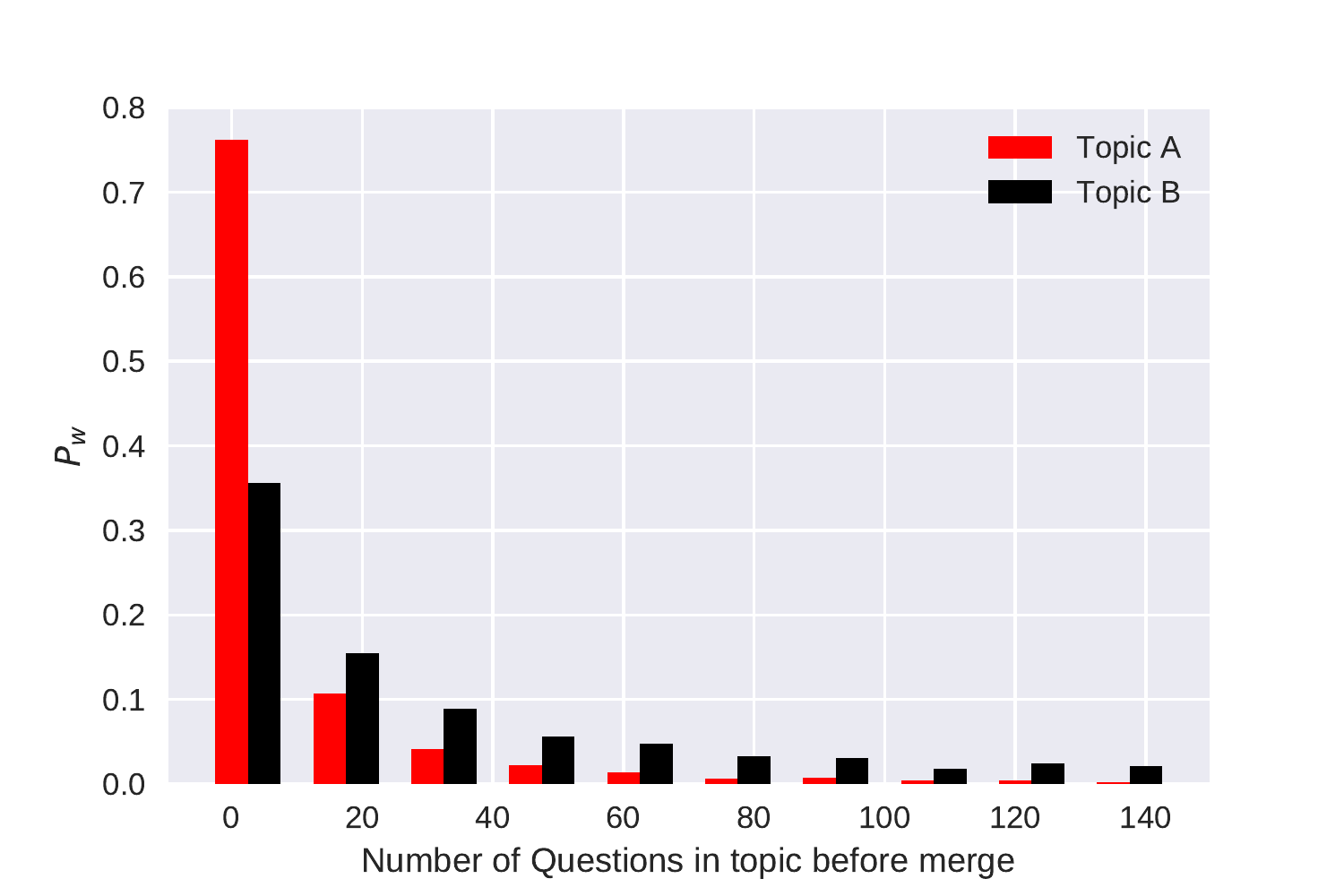}\label{fig:dir_fig3}}
	\subfloat[Number of answers.]{%
		\includegraphics[width=0.24\textwidth]{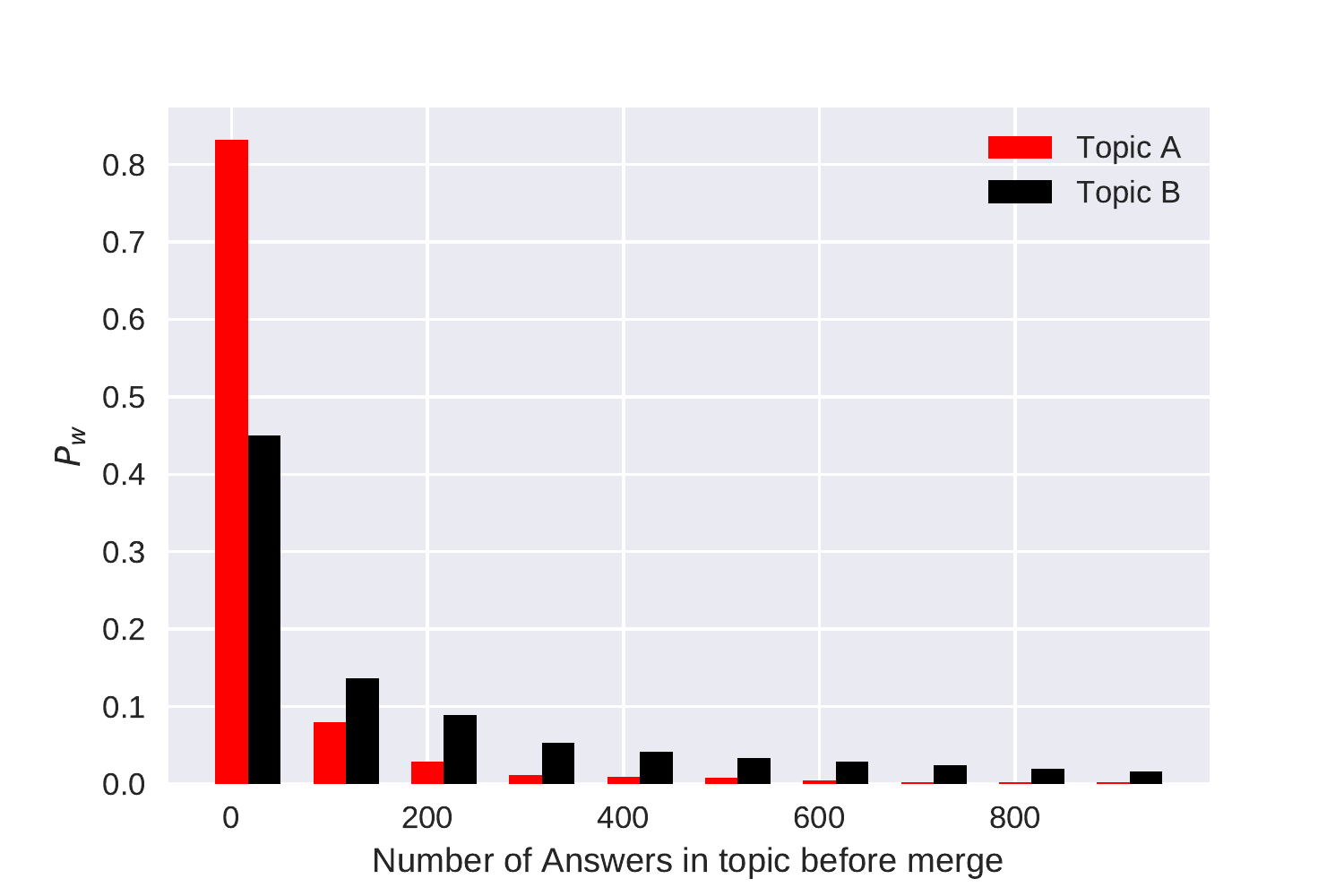}\label{fig:dir_fig4}}\hfill
	\caption{Distribution of various properties influencing the direction of merge (merging of `Topic A' and `Topic B' to `Topic B'). }
\end{figure*}

\noindent\textit{\textbf{Number of characters in the topic name}}:
When deciding the direction of the topic merge, one of the factors to consider is that the name of the topic should be expressive, unambiguous and clear. Therefore, the number of characters plays an important role.
Figure~\ref{fig:dir_fig1} shows the distribution of the number of characters in `Topic A' and `Topic B'. It is evident from the figure that `Topic A' tends to have less number of characters (average 16.5 characters) as compared to `Topic B' (average 23.6 characters). 

\noindent\textit{\textbf{Number of words in the topic name}}:
Similar to the number of characters in the topic name, the number of words in the topic name can also be discriminating. The average number of words in `Topic A' is 2.7 whereas `Topic B' has an average of 3.6 words in the topic name. Figure~\ref{fig:dir_fig2} shows the distribution of the number of words in `Topic A' and `Topic B' which shows similar trend as that of Figure~\ref{fig:dir_fig1}.

\noindent\textit{\textbf{Topic creation date}}:
Another factor involved in deciding the direction of the merge is the time of creation of the topics. 
We compare the date of creation of topic to check if an old topic is being merged into a new one or vice-versa. We find that only 33\% of the times, an older topic gets merged into a newer one.

\noindent\textit{\textbf{Number of questions before merge}}: 
The number of questions in a topic is an indirect measure of how much the topic is used by the community. We calculate the number of questions in both the topics before the merge. Figure~\ref{fig:dir_fig3} shows the distribution of the number of questions in `Topic A' and `Topic B' before they were merged. We can observe that the number of questions in `Topic A' is much lower than that in `Topic B'. On average `Topic A' has 22 questions before the merge whereas `Topic B' has 82 questions before the merge.

\noindent\textit{\textbf{Number of answers before merge}}: 
Similar to the number of questions, number of answers in a topic is also an engagement indicator and might decide the direction of the merge. We calculate the number of answers present in the topics before merging. Figure~\ref{fig:dir_fig4} shows the distribution of the number of answers in `Topic A' and `Topic B'. We can clearly observe that `Topic A' has lesser number of answers as compared to `Topic B'.

\noindent\textbf{Predicting the direction of topic merge}: We use the above discriminating factors in a standard classification framework to predict the direction of topic merge. We use Support Vector Machine (SVM), Random Forest and Naive Bayes classifiers available in Weka Toolkit for classification. We perform a 10-fold cross-validation and report the results in Table~\ref{tab:merge_direction}. We achieve 88.857\% accuracy with high precision and recall. From the table, we can observe that SVM achieves a slightly better performance than the other classifiers.

\begin{table}
	\centering
	\resizebox{8cm}{!}{%
		\begin{tabular}{|c | c | c | c | c | c|} 
			\hline
			Classifier & Accuracy &Precision & Recall & F-Score \\ [0.5ex] 
			\hline\hline &&&&\\[-1.8ex]
			\cellcolor{green}Support Vector Machine & \cellcolor{green}\textbf{88.857} & \cellcolor{green}\textbf{0.828} & \cellcolor{green}\textbf{0.981} & \cellcolor{green}\textbf{0.898} \\
			Random Forest & 88.379 & 0.863 & 0.913 & 0.887 \\
			Naive Bayes & 87.546 & 0.823 & 0.956 & 0.885 \\
			[0.5ex] 
			\hline
		\end{tabular}%
	}
	\caption{Results for predicting the direction of topic merge. For the two-step approach the best five results are only shown.}
	\label{tab:merge_direction}
\end{table}

\subsection{Early prediction of topic merge}
In this section, we discuss the performance of our model to detect duplicate topics which should be merged, at an early stage itself. Note that this could be a direct potential application of our work. We use the best performing method (IF + LR) from the earlier section for our experiments. When a topic is created it takes quite some time for the Quora community to identify if it is a duplicate or not. We find that as per our test dataset, on average it takes 990 days for the user (936 days if we consider the full dataset) to identify a topic as duplicate and merge it. In order to identify how early our model can detect the duplicate topics, we use the following experimental setting.

Let (A, B) be the topic pair, which is known to merge in the future in our test set and let B be the topic that has been created later. We compute the recall of our approach in early detection of this pair. We, therefore, restrict the data to these topics, which are used for computing the features. We take a snapshot every month till both the topics merge. Thus, the first snapshot will have only those questions which have been asked till the first month from the creation of topic B, and so on.
\begin{figure}[tbp]
	\centering
	\includegraphics[width=.95\linewidth]{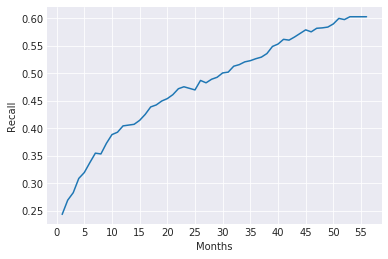} 
	\caption{System recall as a function of time (in months) from the creation of the duplicate topic.}
    \label{fig:early_recall}
\end{figure}

Figure \ref{fig:early_recall} shows the recall of the model on a monthly basis. While the recall keeps improving over time, the result is encouraging since our model is able to detect $\sim 25$\% of the merges in the first month itself, $\sim 40$\% in the first year and $\sim 50$\% in the first 30 months. Thus, the model can potentially be deployed to early detect the topic merges. We stress this early prediction scheme can be directly plugged into Quora enabling efficient merging by both users as well as moderators.

\begin{figure}[htb]
	\centering
	\subfloat[Number of hours spent on Quora by users.]{%
		\includegraphics[width=.23\textwidth]{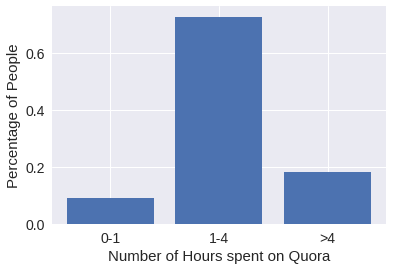} \label{fig:activity}}\hfill
	\subfloat[Familiarity values for the topics assigned to users.]{%
		\includegraphics[width=.23\textwidth]{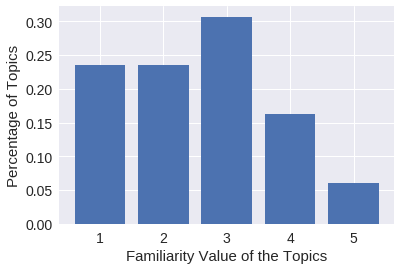} \label{fig:topic_familiarity}}\hfill	
	\caption{Activity and familiarity details of the users.}
	
\end{figure}

\begin{table*}[ht]
	\parbox{.45\linewidth}{
		\centering
		\begin{tabular}{|p{23mm} | p{10mm} |p{12mm} | p{12mm} |} 
			\hline
			Evaluation Methods& Merge & Neighbor & Unmerge \\ [0.5ex] 
			\hline\hline
			Majority Voting & 0.75 & 0.31 & 0.35  \\ 
			Micro-Average & 0.62 & 0.22 & 0.29  \\
			Macro-Average & 0.62 & 0.22 & 0.29 \\ [1ex] 
			\hline
		\end{tabular}%
		\caption{Evaluation results for Question 1.}
		\label{tab:eval_quest1}
	}
	\hfill
	\parbox{.45\linewidth}{
		\centering
		\begin{tabular}{|c | c | c |} 
			\hline
			Evaluation Methods& Question 1 & Question 2 \\ [0.5ex] 
			\hline\hline
			Majority Voting & 0.54 & 0.40  \\ 
			Micro-Average  & 0.33 & 0.49  \\
			Macro-Average & 0.33 & 0.45\\[0.5ex] 
			\hline
		\end{tabular}%
		\caption{Overall human judgment results.}
		\label{tab:heval}}
\end{table*}

\section{Human judgment experiment}
We conduct human judgment experiment to evaluate how our model performs compared to the case where human subjects (regular Quora users) are tasked to predict the merges. We randomly select 180 positive and 180 negative merge cases. The 180 negative cases consists of 90 unmerge and 90 neighborhood pairs. Each pair of topics is annotated by three different users. To understand if humans are able to predict whether two topics should merge or not as well as the direction of merge, we conduct an online survey\footnote{http://tinyurl.com/lgbha5f}. The survey was participated by 45 people (researchers, students and professors) all of whom use Quora. Each participant was asked 24 questions consisting of two parts. Each question had the two topics under consideration along with 3 example questions from each of the topics. The first question asked the participants whether the two topics should merge into a single topic (yes/no type). The second question asked for the direction of merge. A `No' to the first question would enable subjects to skip the second. Apart from these, the participants were asked two additional questions to understand their familiarity / expertise. First, to understand how active they are on Quora, they were asked, ``{\sl How much time (in hours) do you spend on Quora?}''. Second, for each topic pair they annotate, the participants were asked to inform their familiarity with the topic. We presented a scale of 1-5 for the participants to choose from, with the value 1 implying, ``not familiar with the topic'' and a value of 5 implying, ``very familiar with the topic''. In Figure ~\ref{fig:activity}, we show the number of hours spent by the users on Quora. We observe that around 90\% of the participants spend at least one hour on Quora every week. Figure ~\ref{fig:topic_familiarity} shows the percentage of topics with different familiarity values. We observe that very few annotators chose 5 (very familiar). However, most of the annotators were moderately familiar (rating 3) with the topics.

In Table~\ref{tab:eval_quest1}, we report the human evaluation results for each of the three classes separately (Question 1). We observe that humans performed decently with majority voting accuracy of 0.75 for identification of merge classes. For the non-merge cases, humans performed very poorly (worst performance for the neighborhood class). 
These results indicate that the task of merging topics is difficult for Quora users as they seem to be confused by the non-merge cases in which they performed very poorly.
Since the number of participants answering Question 1 for all the questions is 3, the micro-averaging and macro-averaging yields the same result. In Table~\ref{tab:heval}, we show the overall human evaluation results for merge/non-merge identification (Question 1) and identification of direction of merge (Question 2). Overall, we observe that the regular Quora users find it difficult to judge if two given topics should merge or not, and in deciding the direction of merge.

\section{Correspondence analysis}
In this section, we shall compare the outcomes of the automatic prediction framework with the human judgment results. For this purpose, we select the 180 merge and 180 non-merge instances that have been used for the human judgment experiments. This time we train the model on the 2649 (i.e, 2829 - 180) merge cases, 2331 (i.e, 2421 - 90) unmerge cases and 11558 (i.e, 11648 - 90) neighbor instances. For testing, we use 180 merge, 90 unmerge and 90 neighborhood instances\footnote{Since we have a balanced test set here, we only use the supervised classifier and omit the anomaly detection part of the two-step framework} used for the human judgment experiments. We then compare the predicted labels from the model and the human responses decided via majority voting. In Table~\ref{tab:correspondence_analysis}, we present the results of this correspondence analysis. 
The number of concordant cases are higher than the number of discordant cases. The key observation is that 55.61\% of the times both humans and the model agree on the labels; 37.24\% of the times, only the model is able to identify the correct labels while humans fail. We further observe that 46.94\% of the times both human and the model correctly identify the labels; finally only 7.14\% instances were correctly identified by the humans, which were misclassified by the model.

\begin{figure}[ht]
	\centering
    \includegraphics[width=0.47\textwidth]{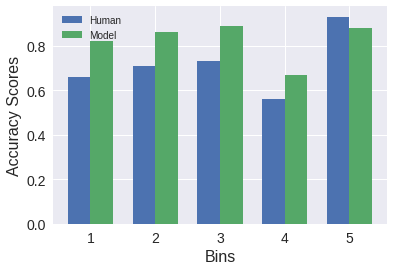}

    \caption{Comparison of accuracies for our model and human judgement}
    \label{fig:bin_wise_accuracy_plot}
\end{figure}

\begin{table}[ht]
	\centering
       \begin{tabular}{||p{5.2cm} | p{1cm}||}
		\hline
		Concordance & 55.61\%\\ \hline
		Discordance & 44.39\%\\ \hline
		Correctly judged by human \& model & 46.94\%\\ \hline
		Wrongly judged by human \& model & 8.67\%\\ \hline
		Correctly judged by only human  & 7.14\%\\ \hline
		Correctly judged by only model & 37.24\% \\ \hline
	\end{tabular}
    \caption{Correspondence analysis}
	\label{tab:correspondence_analysis}
\end{table}

Next, we compare the accuracy of our model against human judgment based on the familiarity of the topics. We use the same model described in the correspondence analysis and vary the test cases based on the familiarity values, as provided by the participants. For every familiarity bin, we calculate the accuracy of the human judgment and the model. Figure ~\ref{fig:bin_wise_accuracy_plot} shows the accuracy comparison between our model and human judgment for each of the bins. Interestingly, we observe the accuracies of humans increases with the familiarity of the topics, with the fourth bin being the only exception. Also, we find that our model performance is better than human annotators for all the bins except the last (very familiar). Thus, only the participants who are experts in the topics (the rarest class) are able to perform better than our model. Even in the case of expert human judgment (Familiarity value 5), our method achieves accuracy that is very close to the domain experts. Considering that a Quora user on average takes 936 days to identify a correct merge, our system can help in finding such possible topic merge quite early. Our system could also act as a filter that could provide the scarce topical experts with possible topic pairs, thus helping in easing the cognitive burden on these experts.

\section{Some insights}

\noindent\textbf{Closely named topics}: Consider the topics `club' and `nightclub'. Our system correctly predicted that these two topics represent two separate concepts and hence should not merge. On the other hand, the human judgment wrongly predicted that these two topics should merge. The topic `club' represents an association of people sharing a common affiliation or interest, whereas the topic `nightclub' represents an entertainment venue and bar which serves alcoholic beverages that usually operates late into the night.

\noindent\textbf{Merges that should have occurred}: Interestingly, our framework was also able to predict some of the merges that should have occurred. For example, consider the topic pair `English-Language-Advice' and `Advice-About-Learning-English'. These were present in our dataset as two different topics and were part of the 1M negative test instances. Our framework predicted that these two topics should merge but as they had not merged till Dec 2016, this was taken as `false positive'. However, these two topics were indeed later merged on 19 Jan 2017. Another such interesting example that we found is `Psycho-2' merged into `Psychology-of-Everyday-Life' on 22 Feb 2017. Automatic enumeration of such correct `false positives' would be an immediate future step.

\noindent\textbf{Enhanced engagement post merge}: To make sure that a topic offers quality content, many edit operations are required to be performed from time to time. We observe that post merging the edit activities in the merged topics have increased. In 70\% of the topics, the volume of edit activity (number of edits/day) increased after the merge. Also in 68\% of the merged topics, the volume of associated questions (number of questions/day) and answers also increased. This suggests that topic merging has an overall positive impact on the quality of the topics and could be key to enhanced user engagement.

\section{Conclusion and future works}
In this paper, we studied the phenomena of competing conventions in Quora topics and proposed a model to predict whether two given topics should merge, as well as the direction of the topic merge. We make use of the Quora topic ontology and question content features to construct the model.

Our two-step approach achieves an F1-score of $\sim 0.71$. We observe that the content features (especially similarity of vector embeddings like word2vec, Doc2vec) and topic ontology together perform significantly well for determining topic merges. In evaluation, the humans were able to predict if two topics should merge or not with an accuracy of 0.54 (compared to $\sim 0.84$ by our framework for the same data). We also propose an early prediction scheme and show that in $\sim 25$\% cases the actual merge pairs can be predicted within a month and in $\sim 40$\% cases within the first year. Users on average require 936 days (as per our dataset) to manually identify such merges. This, we believe, is an effective application that can be directly plugged into the Quora system to enable users/moderators to merge actual topic pairs early in time.  The humans were not able to predict the direction of merge efficiently. Correspondence analysis of the results obtained from the human judgments and the automatic framework shows that our system is able to predict correctly almost all the cases that were predicted by the human judgment. Apart from this, our system was able to correctly predict 37.24\% of the cases that were wrongly predicted by the human judgment. 
Fine-grained analysis based on familiarity of the topics revealed that only the participants very familiar with the topics were able to outperform the proposed model.

An important future direction would be to conduct a detailed error analysis~\cite{derczynski2015analysis,10.1007/978-3-642-40994-3_32} of the false positives obtained from our results which hale be helpful in further improving the model. It would be interesting to automatically enumerate cases from newer time point data, where the merges predicted by the system turned out to be correct later on (but counted as `false positive' now). Such error
analysis can be helpful in further improving the model, and a very precise model can in turn be a very important
contribution to the Quora community.

\bibliography{Main}
\bibliographystyle{aaai}

\end{document}